\documentclass[12pt,twoside]{article}

\pdfoutput=1

\usepackage[utf8]{inputenc}
\usepackage[usenames]{color}
\usepackage{lscape}
\usepackage{amssymb}
\usepackage{rotating}
\usepackage{graphicx}
\usepackage{fancyhdr}
\usepackage{amsmath}
\usepackage{mathrsfs}
\usepackage{ae}
\usepackage{aecompl}
\usepackage{hyperref}

\def\si{s_{\rm in}}
\def\so{s_{\rm out}}
\def\zetai{\zeta^{\rm in}}
\def\zetaisq{(\zeta^{\rm in})^2}

\newlength{\dinwidth}
\newlength{\dinmargin}
\def\oddo{$\,$--$\,$}

\DeclareUnicodeCharacter{00A0}{\nobreakspace}

\begin{document}

\vspace*{1cm}

\begin{center}
\Large\bf {Effects of quantum statistics on relic density\\ of Dark Radiation}
\end{center} 

\vspace*{2mm}

\vspace*{5mm} \noindent
\vskip 0.5cm
\centerline{\bf
Marek Olechowski\footnote[1]{Marek.Olechowski@fuw.edu.pl},
Pawe\l\ Szczerbiak\footnote[2]{Pawel.Szczerbiak@fuw.edu.pl}
}
\vskip 5mm

\centerline{\em Institute of Theoretical Physics,
Faculty of Physics, University of Warsaw} 
\centerline{\em ul.~Pasteura 5, PL--02--093 Warsaw, Poland} 

\vskip 1cm

\centerline{\bf Abstract}
\vskip 3mm

The freeze-out of massless particles is investigated.
The effects due to quantum statistics, Fermi-Dirac or Bose-Einstein,
of all particles
relevant for the process are analyzed. Solutions of appropriate
Boltzmann equation are compared with those obtained using
some popular approximate methods.
As an application of general results the relic density
of dark radiation in Weinberg's Higgs portal model is discussed.

\newpage
\section{Introduction}

In recent decades the cosmic microwave background radiation has been
extensively analyzed, unveiling many crucial facts about the history of the
Universe and its constituents. Although we are convinced that about 27\%
of the total mass-energy fraction of the Universe consists of presumably cold
or warm dark matter, there are still some hints that additional form of
dark radiation (DR) -- called also hot dark matter -- may also exist. 
According to
the recent Planck satellite measurements~\cite{Planck} the effective number
of light neutrino species $N_{\rm eff}$ varies (depending on the effects included)
from $2.99\pm0.20$ to $3.15\pm 0.23\;(1\sigma)$, which indicates that both
the Standard Model (SM) and models with fractional
$\Delta N_{\rm eff}\equiv N_{\rm eff}-N_{\rm eff}^{\rm SM}>0$
are consistent with this result (one fully thermalized neutrino
i.e.~$\Delta N_{\rm eff}=1$ is ruled out at over 3$\sigma$).
On the other hand, the value of the Hubble constant $H_0$ 
obtained from direct observations~\cite{directH0} performed using 
the Hubble Space Telescope i.e.~$73.24\pm1.74\;\rm km\,s^{-1}Mpc^{-1}$ 
is significantly bigger than $67.8\pm0.9\;\rm km\,s^{-1}Mpc^{-1}$ 
inferred from the Planck data. 
Such high value of $H_0$ favors additional contribution
to $\Delta N_{\rm eff}$ (even of order 0.6~\cite{Nunes:2017xon}),
however global fits still prefer the standard $\Lambda$CDM
scenario~\cite{Heavens:2017hkr}.
There is hope that this apparent tension will be clarified in near future,
especially with the advent of the CMB-S4 experiment~\cite{Abazajian:2016yjj}
that will be able to probe $\Delta N_{\rm eff}$ with precision better than 0.03.
So precise experimental results will require more accurate theoretical
treatment of DR freeze-out details than is usually considered in the
literature.

One of the most widely studied scenarios predicting an existence of additional
form of radiation is the Weinberg's Higgs portal model~\cite{Weinberg} in
which a massless Nambu-Goldstone boson of a~broken global U(1) symmetry gives
fractional contribution to $N_{\rm eff}$. It is often assumed that the freeze-out
of that boson takes places just before $\mu^+\mu^-$ annihilation, resulting
in $\Delta N_{\rm eff}\approx 0.39$. However, there are some effects, which we
discuss in this work, that might strongly influence the Goldstone
boson relic density in such case. Moreover, we consider situations
when such boson decouples at different temperature.
Most of our results may be applied also in other models of DR.

The outline of this work is as follows. In section \ref{sec:Boltzmann_eq}
we discuss  
the Boltzmann equation describing the freeze-out of a massless particle
and some approximate methods used in the literature.
Main features of the Weinberg's Higgs portal model 
are given in section \ref{sec:Weinberg_model}.
In section \ref{sec:DR_Weinberg} the relic density of DR in this model
is analyzed in some detail using the general results
presented in section \ref{sec:Boltzmann_eq}.
Section \ref{sec:conclusions} contains our main conclusions.

\section{Boltzmann equation for massless particles
  \label{sec:Boltzmann_eq}}

Boltzmann equation in FLRW metric takes the form
\begin{equation}
\label{BE_general}
E(\partial_t-pH\partial_p)f(p,t)=C_E(p,t)+C_I(p,t)\,,
\end{equation}
where $f(p,t)$ is a distribution function to be found, whereas terms at the 
RHS are collision integrals for elastic ($E$) and inelastic ($I$) processes. 
Elastic collisions are responsible for momentum exchange between particles 
(preserving their relative numbers) and in consequence help to 
maintain kinetic equilibrium. Inelastic collisions are related to 
(co)annihilation processes that influence chemical equilibrium.

Elastic processes are usually stronger than inelastic ones, so we will 
assume here that kinetic equilibrium is maintained. Such assumption is 
not always valid and under certain circumstances may lead to sizable 
differences in relic density calculation.
There are several methods which are applicable for Boltzmann equation
solution without kinetic equilibrium~-- this problem has been extensively
analyzed in the context of neutrino decoupling in the Standard Model.
We can distinguish two major classes 
of relic density calculation for relativistic particles
in such a case: discretization in momentum
space~\cite{discrete,discrete2,discrete3,discrete4} and spectral methods 
based on fixed~\cite{spectral_fixed, spectral_fixed2} or 
dynamical~\cite{spectral_dynamic, spectral_dynamic2} basis of orthogonal 
polynomials. All these approaches allow for high accuracy but at the price of 
lengthy and complicated expressions that are unpractical for qualitative study.
In the present work we are mainly interested in estimation of Bose-Einstein 
(BE)/Fermi-Dirac (FD) statistics corrections to the relic density for 
massless particles. Thus, it suffices in our case to perform analysis
assuming the kinetic equilibrium.

The collision integral for inelastic processes involving particle $\chi$,
$C_I(p_\chi,t)$, with only two-body processes
$\chi a\leftrightarrow bc$ taken into account, may be expressed as
\begin{equation}
\label{CI_general}
\begin{split}
C_I(p_\chi,t)=\frac12\sum_{\{a,b,c\}}\int\prod_{i=a,b,c}&\frac{{\rm d}^3 p_i}{2\pi^3E_i}
(2\pi)^4\delta^{(4)}(p_\chi+p_a-p_b-p_c)\\
\times\sum_{\rm spins}&\left[
|M_{bc\rightarrow \chi a}|^2f_b(p_b,t){f}_c(p_c,t)(1-f_\chi(p_\chi,t))(1-{f}_a(p_a,t))\right.\\
&-\left.|M_{\chi a\rightarrow bc}|^2f_\chi(p_\chi,t){f}_a(p_a,t)(1-f_b(p_b,t))(1-{f}_c(p_c,t))
\right]\,,
\end{split}
\end{equation}
where the sum over $\{a,b,c\}$ corresponds to all allowed processes
$\chi a\leftrightarrow bc$. 
We will calculate the above collision integral exploiting the
method proposed in~\cite{Dolgov}, where the effect of FD statistics was
analyzed in the context of relic density calculation for relativistic
fermions ($0\not=m\ll T_f$) staying in kinetic equilibrium. Below we will
focus on annihilation of massless particles (bosons and fermions) including
full statistics for both initial and final states.
The results will be used in section~\ref{sec:DR_Weinberg} to analyze the
Goldstone bosons freeze-out in the Weinberg's Higgs portal model.

Let us consider annihilation process of the form $\bar{\chi}\chi\to\bar{N}N$,
where $N$ stays in kinetic and chemical equilibrium during $\chi$ freeze-out.
For definiteness, we will focus on one process of this kind (generalization
is straightforward). Distribution functions for $\chi$ and $N$ may be written
in the form
\begin{equation}
  f_\chi(p,t)\simeq\left(e^{E/T+z}\mp1\right)^{-1}\equiv\left(e^{xy+z}
  + \si \right)^{-1}\,,
\end{equation}
\begin{equation}
f_N(p,t)\simeq\left(e^{E_N/T} + \so \right)^{-1}\,,
\end{equation}
where $z=z(t)$ is the so-called chemical pseudopotential (equal for $\chi$
and $\bar\chi$)~\cite{pseudo_def}.
We also defined $x\equiv m/T$ and $y\equiv E/m$ (for one massive
annihilation product $N$ it is convenient to choose $m\equiv m_N\not=0$).
The statistical factors $\si$ and $\so$ equal $+1$ and $-1$ for fermions
and bosons, respectively~-- by setting these factors to zero one obtains 
the Maxwell-Boltzmann (MB) approximation.
Using the above definitions and integrating expression~\eqref{CI_general} over $\chi$ momenta one may rewrite eq.~\eqref{BE_general}
in the following form~\cite{Dolgov}
\begin{equation}
\label{BE_z}
\frac{{\rm d}z}{{\rm d}x}\simeq
\left(A(z,x)\frac{x}{T}\frac{{\rm d}T}{{\rm d}t}\right)^{-1}\left(S_I(z,x)-B(z,x)\right)\,,
\end{equation}
where
\begin{equation}
\begin{split}
A(z,x)&\equiv
\frac{g}{2\pi^2}m^3e^z J_2(z,x)\,,\\
B(z,x)&\equiv
\frac{g}{2\pi^2}m^3e^zx J_3(z,x)H(x)
\left(1-\frac{1}{1-\frac{x}{3}\frac{g_{*s}'(x)}{g_{*s}(x)}}\right)\,,\\
S_I(z,x)&\equiv
\int\frac{{\rm d}^3p}{(2\pi)^3}\frac{1}{E}C_I(p,t)\,,
\end{split}
\end{equation}
$g$ is the number of $\chi$ degrees of freedom,
the Hubble parameter is given by
\begin{equation}
H(x)=
\sqrt{\frac{4\pi^3}{45}}\sqrt{g_{*}(x)}\frac{m^2}{M_{\rm Pl}}\frac{1}{x^2}\,,
\end{equation}
and\footnote{
  Note the difference with respect to $J_n$ function defined in eq.~(10)
  in~\cite{Dolgov}.}
\begin{equation}
\label{J}
J_n(z,x)\equiv\int_0^\infty {\rm d}y\,y^n\frac{e^{xy}}{(e^{xy+z}+ \si)^2}\;.
\end{equation}
The most elaborate part is the calculation of the collision integral $S_I$
which, after using four-momentum conservation, is 5-dimensional. We use
the following convenient set of variables: energies of the incoming particles,
$E_1$ and $E_2$, one of the outgoing particles' energy, $E_3$, the angle
between momenta of the incoming particles, $\theta$, and the acoplanarity angle
$\phi$. Defining dimensionless parameters $u\equiv E_1/m$, $v\equiv E_2/m$
and $t\equiv E_3/m$ one can write
\begin{equation}
\label{S_I_general}
S_I(z,x)=
\frac{m^4}{512\pi^6}\left(e^{2z}-1\right)
\int\mathscr{D}\Phi\int_0^{2\pi}{\rm d}\phi
\sum_{\rm spins}|M_{\chi\bar{\chi}\to N\bar{N}}(u,v,t,\cos\theta,\phi)|^2\,,
\end{equation}
where
\begin{equation}
\label{Dphi_general}
\begin{split}
\mathscr{D}\Phi&\equiv
\int_0^\infty {\rm d}u\int_0^\infty {\rm d}v\int_{-1}^1{\rm d}\cos\theta\frac{uv\,e^{x(u+v)}}{\kappa(u,v,\theta)}\int_{t_-}^{t_+}{\rm d}t\\
&\times\frac{1}{
\left(e^{xu+z}+ \si\right)
\left(e^{xv+z}+ \si\right)
\left(e^{xt}+ \so\right)
\left(e^{x(u+v-t)}+ \so\right)}
\end{split}
\end{equation}
and we introduced the following functions of the parameters: 
$\kappa(u,v,\theta)\equiv(u^2+v^2+2\,uv\cos\theta)^{1/2}$,
$V(u,v,\theta)\equiv(1-4m^2/s)^{1/2}$, 
$t_\pm\equiv\frac12\left[u+v\pm\kappa(u,v,\theta)V(u,v,\theta)\right]$.
If (which is often the case in leading approximation)
$M_{\chi\bar{\chi}\to N\bar{N}}$ depends only on $s$
(given here by $s=2uv\,m^2(1-\cos\theta)$), 
introducing new combinations of the parameters 
$p=x(u+v)$ and $q=x\sqrt{u^2+v^2+2uv\cos\theta}$,
we can reduce the above integral to
\begin{equation}
\label{Dphi_int_cos}
\begin{split}
\mathscr{D}\Phi&=
\frac{1}{x^4}
\int_{2x}^\infty {\rm d}p\int_0^{\sqrt{p^2-4x^2}} {\rm d}q\;
\frac{1}{(|\so|-e^{-p})(e^{p+2z} - |\so|)}\\
&\times
\ln\left[
\frac{\cosh\left(\frac12(p+q)+z\right)+ \si}
{\cosh\left(\frac12(p-q)+z\right)+ \si}
\right]
\ln\left[
\frac{\cosh\left(\frac12\big(p+qV(p,q)\big)\right)+ \so}
{\cosh\left(\frac12\big(p-qV(p,q)\big)\right)+ \so}
\right]\,,
\end{split}
\end{equation}
where now $V=\left(1-\frac{4x^2}{p^2-q^2}\right)^{1/2}$ 
and $s=\frac{p^2-q^2}{x^2}m^2$.
In such a case the integration over $\phi$
in eq.~\eqref{S_I_general} is trivial and the expression simplifies.
Then, the Boltzmann equation~\eqref{BE_z} may be written in the following
integro-differential form 
\begin{equation}
\label{BE_zx}
\frac{{\rm d}z}{{\rm d}x}=-\frac{x}{J_2(z,x)}
\left(
\frac{1}{3}\frac{g'_{*s}(x)}{g_{*s}(x)}J_3(z,x)+
\frac{\sqrt{45}}{256\pi^{11/2}}\frac{M_{\rm Pl}}{m}
\frac{\sinh z}{g\sqrt{g_{*}(x)}}
\left(1-\frac{x}{3}\frac{g'_{*s}(x)}{g_{*s}(x)}\right)\tilde{S}_I(z,x)
\right)\,.
\end{equation}
Note that we introduced dimensionless (in contrast to $S_I(z,x)$) collision
integral (see eq.~\eqref{Dphi_int_cos})
\begin{equation}
\label{S_I_tilde}
\tilde{S}_I(z,x)=
2\pi\int\mathscr{D}\Phi\sum_{\rm spins}|M_{\chi\bar{\chi}\to N\bar{N}}(s)|^2\,.
\end{equation}
The number density of $\chi$ ($=\bar{\chi}$) may be easily found after
integration over the whole energy
range\footnote{
  If $\chi$ and $\bar{\chi}$ are distinguishable one should consider
  $n = n_\chi + n_{\bar{\chi}}$ instead.}
\begin{equation}
\label{n}
n(x)\equiv n_\chi(x)=g\frac{m^3}{2\pi^2}\int_0^\infty\frac{y^2{\rm d}y}{e^{xy+z}+ \si}\;.
\end{equation}
%

\subsection{MB approximation and limited inclusion of BE/FD statistics
  \label{subs:app}}

Let us consider the particle number density normalized by the entropy
density: $Y(x)\equiv\frac{n(x)}{s(x)}$. In chemical and kinetic
equilibrium (subscript eq) we just have
(using eq.~\eqref{n} with $z=0$)
\begin{equation}
\label{Yeq}
Y_{\rm eq}(x)=g\,\frac{45}{2\pi^4}\frac{\zetai}{g_{*s}(x)}\;,
\end{equation}
where
\begin{equation}
\label{zeta_mp}
\zetai\equiv\zeta(3)
\begin{cases}
1 & \text{BE}\\
3/4 & \text{FD}
\end{cases}\,.
\end{equation}
For the Maxwell-Boltzmann statistics one can write
\begin{equation}
\label{y_and_Yeq_MB}
Y(x)=e^{-z}Y_{\rm eq}(x)\,.
\end{equation}
Then, the Boltzmann equation~\eqref{BE_zx} simplifies to
(we use $\si=\so=0$ in eqs.~\eqref{J} and \eqref{Dphi_general})
\begin{equation}
\label{z_MB}
\frac{{\rm d}z}{{\rm d}x}=\
-\frac{g'_{*s}(x)}{g_{*s}(x)}-
\frac{\sqrt{45}}{\pi^{7/2}}\frac{M_{\rm Pl}m}{x^2}\frac{\sinh z}{g\sqrt{g_{*}(x)}}
\left(1-\frac{x}{3}\frac{g_{*s}'(x)}{g_{*s}(\tilde{x})}\right)
\langle\sigma v\rangle_{\rm MB}\;.
\end{equation}
One can also rewrite it in the Lee-Weinberg form
\begin{equation}
\label{Y_MB}
Y'(x)=-\sqrt{\frac{\pi}{45}}
\frac{g_{*s}(x)}{\sqrt{g_{*}(x)}}
  \frac{M_{\rm Pl}m}{x^2}\left(Y^2(x)-Y^2_{\rm eq}(x)\right)\langle\sigma v\rangle_{\rm MB}\frac{1}{\zetai}\;,
\end{equation}
where $v$ is the M\"{o}ller velocity for the incoming particles and 
\begin{equation}
\label{sigmav_MB}
\langle\sigma v\rangle_{\rm MB}=
\frac{1}{512\pi}\frac{x^5}{m^5}
\int_{4m^2}^\infty {\rm d}s\,\sqrt{s}\,
\sqrt{1-\frac{4m^2}{s}}\,\sum_{\rm spins}|M(s)|^2{\rm K}_1
\left(\frac{x\sqrt{s}}{m}\right)\,.
\end{equation}
Please note the presence of additional factor $1/\zetai$ in the RHS of
eq.~\eqref{Y_MB} as compared to the standard form~\cite{GondoloGelmini}.
It comes from the fact that $Y_{\rm eq}$ contains information about the incoming
particles statistics whereas eq.~\eqref{BE_zx}, after putting everywhere
$\so=\si=0$, does not. Multiplying $\tilde{S}_I(z,x)$ by $\zetai$ one can
obtain the familiar Lee-Weinberg equation. Thus, we have two options:
\begin{enumerate}
\item Use eqs.~\eqref{Yeq} and \eqref{Y_MB} with $\zetai=1$
  (\textit{pure} MB approximation).
\item Use eqs. \eqref{Yeq} and \eqref{Y_MB} with $\zetai$ given by
  eq.~\eqref{zeta_mp} in the former but $\zetai = 1$ in the latter (we call it \textit{fractional} inclusion of BE/FD statistics for
  incoming particles and denote as fBE/fFD).
\end{enumerate}
One can also include BE/FD statistics (again, only for incoming particles)
in thermally averaged cross section
\begin{equation}
\label{sigmav_pBEpFD}
\begin{split}
\langle\sigma v\rangle_{\rm p}&=
\frac{1}{512\pi\zetaisq}\frac{x^5}{m^5}
\int_{4m^2}^\infty {\rm d}s\,\sqrt{1-\frac{4m^2}{s}}\,\sum_{\rm spins}|M(s)|^2\\
&\times\int_{\sqrt{s}}^\infty {\rm d}E_+\frac{e^{-\frac{x}{2m}E_+}}{\sinh\left(\frac{x}{2m}E_+\right)}
\ln\left[
\frac{{\rm fh}\left(\frac{x}{4m}\left(E_++\sqrt{E_+^2-s}\right)\right)}
{{\rm fh}\left(\frac{x}{4m}\left(E_+-\sqrt{E_+^2-s}\right)\right)}
\right]\,,
\end{split}
\end{equation}
where ${\rm fh}$ is a hyperbolic function depending on the statistics
of incoming particles: ${\rm fh}\equiv\sinh$ ($\cosh$) for bosons
(fermions) and index p stands for partial inclusion of statistics.
Thus, we consider another approximate method:
\begin{enumerate}
\item[3.]
  The same rules as in the point 2.~above but with
  $\langle\sigma v\rangle_{\rm p}$ (eq.~\eqref{sigmav_pBEpFD}) instead of
  $\langle\sigma v\rangle_{\rm MB}$ (eq.~\eqref{sigmav_MB})  
  (we call it \textit{partial} inclusion of BE/FD statistics for
  incoming particles and denote as pBE/pFD).
\end{enumerate}
The main difference between eqs.~\eqref{Dphi_int_cos} and \eqref{sigmav_pBEpFD}
is that in the latter case the expression does not depend on $Y$
(or equivalently on $z$), which significantly simplifies calculations.
Note also that none of the above approximations include any effect of
the outgoing particles statistics.

\section{Weinberg's Higgs portal model}
\label{sec:Weinberg_model}

For easy reference we remind here the main features of the model
proposed in \cite{Weinberg} (using notation mainly from
\cite{Weinberg_notation}). In addition to the SM fields it contains
a dark sector consisting of a complex scalar $\phi$ and a Dirac
fermion $\psi$. Both new fields are charged only under a global
$U(1)_{\rm dark}$ symmetry with charges:
$Q_{\rm dark}(\psi)=1$, $Q_{\rm dark}(\phi)=2$.
A non-zero VEV of $\phi$ spontaneously breaks this symmetry to
the discrete $\mathbb Z_2$ parity. The Goldstone boson associated
with this symmetry breaking contributes to DR while the lighter 
dark fermion ($\psi$ gives two Majorana fermions with different masses)
is stable and plays the role of cold dark matter (CDM).

We will concentrate here on the scalar sector which is the most important
one for our analysis. The corresponding part of the Lagrangian
density is given by
\begin{equation}
\label{W_LH}
\begin{split}
\mathcal{L}_{H,\phi} =& \left(D_\mu H\right)^\dagger\left(D^\mu H\right)+
\mu_H^2H^\dagger H-
\lambda_H(H^\dagger H)^2\\
+ & \partial_\mu\phi^*\partial^\mu\phi + \mu_\phi^2(\phi^*\phi)^2
-\lambda_\phi(\phi^*\phi)^2
-\kappa(H^\dagger H)(\phi^*\phi)\;,
\end{split}
\end{equation}
where $H$ is the Standard Model Higgs doublet.
After the spontaneous symmetry breaking the scalar fields may be written as
\begin{equation}
\label{W_fields}
H=\begin{pmatrix}
G^+\\ v_H + \frac{\tilde{h}+iG^0}{\sqrt{2}}
\end{pmatrix}\;,\quad\quad
\phi = v_\phi + \frac{\tilde{\rho}+i\sigma}{\sqrt{2}}\;,
\end{equation}
where $\sigma$ is the Goldstone boson of the broken
$U(1)_{\rm dark}$ and contributes to DR. Two neutral scalars $\tilde{h}$
and $\tilde{\rho}$ mix, forming two mass eigenstates ($h$ and $\rho$):
\begin{equation}
\label{h+rho}
h=\tilde{h}\cos\theta+\tilde{\rho}\sin\theta
\;,\quad\quad
\rho = \tilde{\rho}\cos\theta-\tilde{h}\sin\theta\;,
\end{equation}
with masses
\begin{equation}
\label{W_mh2}
m_h^2 = 4\lambda_H v_H^2\cos^2\theta + 4\lambda_\phi v_\phi^2\sin^2\theta + 2\kappa\, v_H v_\phi\sin2\theta\,,
\end{equation}
\begin{equation}
\label{W_mrho2}
m_\rho^2 = 4\lambda_\rho v_\rho^2\cos^2\theta + 4\lambda_H v_H^2\sin^2\theta - 2\kappa\, v_H v_\phi\sin2\theta\,,
\end{equation}
and the mixing angle given by the condition
\begin{equation}
\tan2\theta=\frac{\kappa\, v_Hv_\phi}{\lambda_H v_H^2 - \lambda_\phi v_\phi^2}\;.
\end{equation}
The scalar potential in \eqref{W_LH} depends on five parameters. They must
satisfy two conditions in order to give the correct values of $v_H$
and the mass of the Higgs particle (identified with $h$). As the three
parameters, describing the remaining freedom in the scalar Lagrangian
\eqref{W_LH}, we choose two couplings, $\lambda_\phi$ and $\kappa$,
and the mass of the non-SM-like scalar, $m_\rho$.
Other parameters present in the mass formulas \eqref{W_mh2} and \eqref{W_mrho2}
may be expressed as
\begin{equation}
\label{W_sinT}
\sin\theta\simeq 
\frac{\kappa\, m_\rho v_H}{\left[\lambda_\phi(m_h^2-m_\rho^2)^2 - \kappa^2v_H^2(m_h^2-m_\rho^2)\right]^{1/2}}
\approx 
\frac{\kappa}{\lambda_\phi^{1/2}}\left(\frac{m_\rho}{90\;\,\rm GeV}\right)\;,
\end{equation}
\begin{equation}
\label{W_lamH}
\lambda_H = \frac{m_h^2\cos^2\theta + m_\rho^2\sin^2\theta}{4v_H^2}\approx 0.13\cos^2\theta\;,
\end{equation}
\begin{equation}
\label{W_vPhi}
v_\phi^2 = \frac{m_\rho^2\cos^2\theta + m_h^2\sin^2\theta}{4\lambda_\phi}\approx
m_\rho^2\left(\frac{\cos\theta}{2\lambda_\phi^{1/2}}\right)^2\;,
\end{equation}
where the approximate equalities are valid if the scalar $\rho$ is
relatively light and does not mix strongly with the SM-like Higgs i.e.\
$m^2_\rho\ll m^2_h$, $\sin^2\theta\ll 1$.

Let us now estimate phenomenologically interesting range of values for 
$\kappa$ and $\lambda_\phi$ that will be used in our numerical scan.
The contribution from the dark particles to the invisible Higgs decay width
may be approximated as
\begin{equation}
\label{W_Gamma_h_inv}
\Delta\Gamma_{h,{\rm inv}}
\geq
\Gamma(h\to\sigma\sigma)+
\Gamma(h\to\rho\rho)
\approx
\frac{\sin^2\theta}{32\pi}\frac{m_h^3}{v_\phi^2}
\approx
\frac{\kappa^2}{8\pi\cos^2\theta}\frac{v_H^2}{m_h}\,,
\end{equation}
where we conservatively neglected the Higgs decays into dark fermions.
The LHC constraint on the invisible Higgs decays i.e.
\begin{equation}
\label{W_Gamma_h_inv_LHC}
\Delta\Gamma_{h,{\rm inv}}<
\frac{B_{\rm inv}\cos^2\theta}{1 - B_{\rm inv}}\Gamma_{h,\rm SM}\,,
\end{equation}
where $B_{\rm inv}\lesssim 24\%$~\cite{W_LHC_inv} and
$\Gamma_{h,\rm SM}\simeq 4$~MeV, 
may be used together with eq.~\eqref{W_Gamma_h_inv}
to obtain the following upper bound on the portal coupling:
\begin{equation}
\label{W_kappa_LHC}
\kappa\lesssim 0.01\,.
\end{equation}
Bounds on the singlet scalar self-coupling, $\lambda_\phi$, are much weaker. They follow from the requirement
of perturbativity to some high energy scale and may be obtained
from the RG equations~\cite{W_RGE}:
\begin{equation}
\label{W_RGE}
\begin{split}
16\pi^2\frac{{\rm d}\lambda_\phi}{{\rm d}t} &= 10\lambda_\phi^2 + \kappa^2 \,,\\
16\pi^2\frac{{\rm d}\kappa}{{\rm d}t} &
=
\kappa\left(2\kappa + 4\lambda_\phi + 6\lambda_H + 3 y_t^2
-\frac94g_2^2 - \frac34g_1^2\right)\,,
\end{split}
\end{equation}
where $t\equiv Q^2/Q_0^2$. The scalar couplings stay perturbative up to the
GUT energy scale if at low energy $\lambda_\phi\lesssim0.25$. On the other hand,
if $\lambda_\phi=1$ at the weak scale, it becomes non-perturbative already at scale of order 100~TeV. Both values i.e.\ $\lambda_\phi = 0.25,\,1$ will be our reference points in numerical scans in the next section.

\section{Dark Radiation in Weinberg's Higgs portal model 
\label{sec:DR_Weinberg}}

In this section we will compare solutions of the Boltzmann equation~\eqref{BE_zx} with the instantaneous freeze-out approximation as well as with three approximate methods 
introduced in section \ref{subs:app}. 
The model of DR proposed by Steven Weinberg \cite{Weinberg} 
will be used as a testing ground for this purpose.

In order to calculate the relic abundance of the Goldstone boson $\sigma$ 
one needs to know its cross section for the annihilation into SM 
particles. In the original paper \cite{Weinberg} it was assumed that 
$\sigma$ decouples at temperature (just) above the muon mass.
Then, in some analyses (e.g.~\cite{Weinberg_notation, 1312.2547})
only the annihilation of $\sigma$ into muons was taken into account. 
However, pions are not much heavier than muons and their 
contribution should also be included. It was
shown~\cite{Voloshin, higgs_hunter} that the effect from pions may be
a few times stronger than the one from muons (in the context of Weinberg's
Higgs portal model it was analyzed e.g.~in~\cite{pions}).
The squares of matrix elements for the relevant processes with $\rho$
exchanged in the s channel may be approximated 
as\footnote{
  We neglected the mass difference between the charged and neutral pions and
  summed effectively their contributions at the level of amplitudes instead
  of cross sections.
  }
\begin{equation}
\label{W_M2_mions}
\sum_{\rm spins}|M_{\sigma\sigma\to\mu^+\mu^-}(s)|^2 \approx 2\kappa^2s^2
\frac{m_{\mu^\pm}^2(s-4m_{\mu^\pm}^2)}
{\left[(s-m_\rho^2)^2 + \Gamma_\rho^2m_\rho^2\right]
\left[(s-m_h^2)^2 + \Gamma_h^2m_h^2\right]}\,,
\end{equation}
\begin{equation}
\label{W_M2_pions}
\sum_{\rm spins}|M_{\sigma\sigma\to\pi\pi}(s)|^2 \approx 2\kappa^2s^2
\frac{\frac{1}{27}(s+\frac{11}{2}m_\pi^2)^2}
{\left[(s-m_\rho^2)^2 + \Gamma_\rho^2m_\rho^2\right]
\left[(s-m_h^2)^2 + \Gamma_h^2m_h^2\right]}\,.
\end{equation}
We checked that the effects from $\tau^\pm$, gluons and free quarks
become important at temperatures $T\gtrsim 0.4\div 0.5$~GeV.
This suggests that the Goldstone bosons stay in equilibrium
at temperatures above $\Lambda_{\rm QCD}$. Thus, we assume in our
analysis that $\sigma$ particles indeed are in equilibrium for $T>0.4$~GeV.

It occurs that from the phenomenological point of view the most interesting
region of the parameter space is that close to the resonance
i.e.~$s\sim m_\rho^2$. If the resonance is narrow, i.e.~when $\left({\Gamma_\rho}/{m_\rho}\right)^2\ll 1$, one may use the approximation
\begin{equation}
\label{narrow_res}
\frac{1}{\pi}\frac{m_\rho\Gamma_\rho}{(s-m_\rho^2)^2 + \Gamma_\rho^2m_\rho^2}
\longrightarrow \delta(s-m_\rho^2)\,.
\end{equation}
We assume for simplicity that the CDM fermion is not very light
and fulfills the condition $m_{\rm CDM}>m_\rho/2$. Then the width of $\rho$
is dominated by its decays into Goldstone bosons pairs, giving
\begin{equation}
\label{W_Gamma_rho}
\Gamma_\rho\approx\Gamma(\rho\to\sigma\sigma)=
\frac{m_\rho^3\cos^2\theta}{64\pi v_\phi^2}\approx
m_\rho\left(\frac{\lambda_\phi}{16\pi}\right)\,.
\end{equation}
Because $\left({\Gamma_\rho}/{m_\rho}\right)^2=(\lambda_\phi/16\pi^2)^2\ll 1$
approximation \eqref{narrow_res} may be used even for $\lambda_\phi$ as big as 1.
We do not consider larger values of $\lambda_\phi$ because, as explained
below eq.~\eqref{W_RGE}, they would lead to a cut-off scale below 100~TeV.
Thus, we may apply the narrow-resonance approximation
\eqref{narrow_res} and rewrite the matrix elements \eqref{W_M2_mions}
and \eqref{W_M2_pions} in the following form
\begin{equation}
\label{W_M2_mions_delta}
\sum_{\rm spins}|M_{\sigma\sigma\to\mu^+\mu^-}|^2 = 2\pi\kappa^2m_\rho^3\,\delta(s-m_\rho^2)
\frac{m_{\mu^\pm}^2(m_\rho^2-4m_{\mu^\pm}^2)}
{\Gamma_\rho\left[(m_\rho^2-m_h^2)^2 + \Gamma_h^2m_h^2\right]}\,,
\end{equation}
\begin{equation}
\label{W_M2_pions_delta}
\sum_{\rm spins}|M_{\sigma\sigma\to\pi\pi}|^2 = 2\pi\kappa^2m_\rho^3\,\delta(s-m_\rho^2)
\frac{\frac{1}{27}(m_\rho^2+\frac{11}{2}m_\pi^2)^2}
{\Gamma_\rho\left[(m_\rho^2-m_h^2)^2 + \Gamma_h^2m_h^2\right]}\,.
\end{equation}
Using this approximation we can also simplify integrals~\eqref{Dphi_int_cos},
\eqref{sigmav_MB} and \eqref{sigmav_pBEpFD},
obtaining
\begin{equation}
\label{Dphi_int_cos_Wdelta}
\begin{split}
\mathscr{D}\Phi&=
\frac{1}{2x^2}
\int_{1}^\infty {\rm d}y\frac{1}{\sqrt{y^2-1}}\;
\frac{1}{(|\so|-e^{-p})(e^{p+2z}-|\so|)}\\
&\times
\ln\left[
\frac{\cosh\left(\frac12(p+q)+z\right)+ \si}
{\cosh\left(\frac12(p-q)+z\right)+ \si}
\right]
\ln\left[
\frac{\cosh\left(\frac12\big(p+qV\big)\right)+ \so}
{\cosh\left(\frac12\big(p-qV\big)\right)+ \so}
\right]\,,
\end{split}
\end{equation}
\begin{equation}
\label{sigmav_MB_Wdelta}
\langle\sigma v\rangle_{\rm MB}=
\frac{\sum_{\rm spins}|M|^2}{512\pi}\frac{x^5}{m_{\mu^\pm}^2}
\sqrt{\frac{m_\rho^2}{m_{\mu^\pm}^2}-4}
\;{\rm K}_1\left(\frac{x\,m_\rho}{m_{\mu^\pm}}\right)\,,
\end{equation}
\begin{equation}
\label{sigmav_pBEpFD_Wdelta}
\begin{split}
\langle\sigma v\rangle_{\rm p}&=
\frac{\sum_{\rm spins}|M|^2}{512\pi}\frac{x^5}{m_{\mu^\pm}^3}\frac{1}{\zetaisq}
\sqrt{1-\frac{4m_{\mu^\pm}^2}{m_\rho^2}}
\\
&\times\int_{m_\rho}^\infty {\rm d}E_+\frac{e^{-\frac{x}{2m_{\mu^\pm}}E_+}}
  {\sinh\left(\frac{x}{2m_{\mu^\pm}}E_+\right)}
\ln\left[
  \frac{{\rm fh}\left(\frac{x}{4m_{\mu^\pm}}
    \left(E_++\sqrt{E_+^2-m_\rho^2}\right)\right)}
{{\rm fh}\left(\frac{x}{4m_{\mu^\pm}}\left(E_+-\sqrt{E_+^2-m_\rho^2}\right)\right)}
\right]\,,
\end{split}
\end{equation}
which considerably simplify numerical
calculations.\footnote{
Equation \eqref{Dphi_int_cos_Wdelta} follows from \eqref{Dphi_int_cos}
after changing variables and performing  one integration, using
\eqref{narrow_res}.
The functions appearing in the integral~\eqref{Dphi_int_cos_Wdelta}
depend on the integration variable $y$ as follows:
$p=x(m_\rho/m_{\mu^\pm})y$, $q=x(m_\rho/m_{\mu^\pm})\sqrt{y^2-1}$ and
$V=\sqrt{1-4(m_{\mu^\pm}/m_\rho)^2}$.}

Before presenting the results of our numerical analysis let us remind
the definition of the effective number of neutrino species, $N_{\rm eff}$.
It is usually defined by the formula
\begin{equation}
\label{Neff_def}
\rho_R = \left[ 1 + \left(\frac{\rho_\nu^{\rm ifo}}{\rho_\gamma^{\rm ifo}}\right)
  N_{\rm eff}\right]\rho_\gamma
= 
\left[ 1 + \frac78\left(\frac{4}{11}\right)^{4/3} N_{\rm eff}\right]\rho_\gamma \,,
\end{equation}
where $\rho_R$, $\rho_\gamma$ and $\rho_\nu$ correspond to the present total,
photons and neutrinos radiation energy density, respectively.
Index ifo denotes instantaneous freeze-out approximation.
In general we can write
\begin{equation}
\label{rhoR_def}
\rho_R = \rho_\gamma + \sum_{i=1}^3\rho_{\nu_i} + \rho_X \,,
\end{equation}
where $\rho_X$ is an additional form of radiation existing in certain
extensions of the Standard Model. In the case of Weinberg's Higgs portal model: $X=\sigma$.
$N_{\rm eff}$ is normalized in such a way that using the instantaneous
freeze-out approximation
for neutrino decoupling and putting $\rho_X=0$ we get $N_{\rm eff}=3$.
In the Standard Model due to the lack of perfect kinetic equilibrium during
neutrino decoupling one gets $N_{\rm eff}^{\rm SM}\approx 3.046$~\cite{Mangano_SM}.
For simplicity, we will define additional effective number of neutrino species as 
\begin{equation}
\label{deltaNeff_def}
\Delta N_{\rm eff}\equiv N_{\rm eff} - N_{\rm eff}^{\rm SM} 
\,,
\end{equation}
which for the Goldstone boson $\sigma$ 
(additional factor $\frac78$ due to different statistics and number
of degrees of freedom as compared to $\nu$) gives
\begin{equation}
\label{deltaNeff_def_exp}
\Delta N_{\rm eff} =
\frac47\left(\frac{T_\gamma^{\rm ifo}}{T_\nu^{\rm ifo}}\right)^4\frac{\rho_\sigma}{\rho_\gamma}
=
\left.\frac47\frac{\rho_\sigma}{\rho_\gamma}\right|_{T=T_{\rm end}}
=
\left.\frac{4}{7}\left(\frac{Y}{Y_{\rm eq}}\right)^{\frac{4}{3}}\right|_{x=x_{\rm end}}
\,,
\end{equation}
where $x_{\rm end}\approx20$ corresponds to $T_{\rm end}\approx 5$~MeV
(temperature before neutrino decoupling when the freeze-out process
of $\sigma$ is completed).

\subsection{Instantaneous freeze-out approximation
  \label{subs:ifo_app}}

Instantaneous freeze-out approximation is one of the simplest methods for
relic density estimation. By definition, freeze-out takes place at such $x_f$
for which the freeze-out parameter $\eta(x)$, being the ratio between
the interaction rate $\Gamma$ and the Hubble parameter $H$, equals one:
\begin{equation}
  \eta(x) =
  \left.\frac{\Gamma}{H}\right|_x
  =
  \left.\frac{n\langle\sigma v\rangle}{H}\right|_x\,,\quad\quad
\eta(x_f)=1\,.
\label{freeze_app}
\end{equation}
At temperatures below $T_f$ ($x>x_f$) the yield of considered particles
remains constant i.e.\ $Y(x)=Y^{\rm eq}(x_f)$. 
For $\langle\sigma v\rangle = \langle\sigma v\rangle_{\rm MB}$
(see eq.~\eqref{sigmav_MB}) with matrix elements squared as
in~\eqref{W_M2_mions_delta}\oddo\eqref{W_M2_pions_delta}
and $\Gamma_\rho$ given by~\eqref{W_Gamma_rho}, one gets the following
condition for $x_f$
\begin{figure}[t!]
\center
\includegraphics[width=0.49\textwidth]{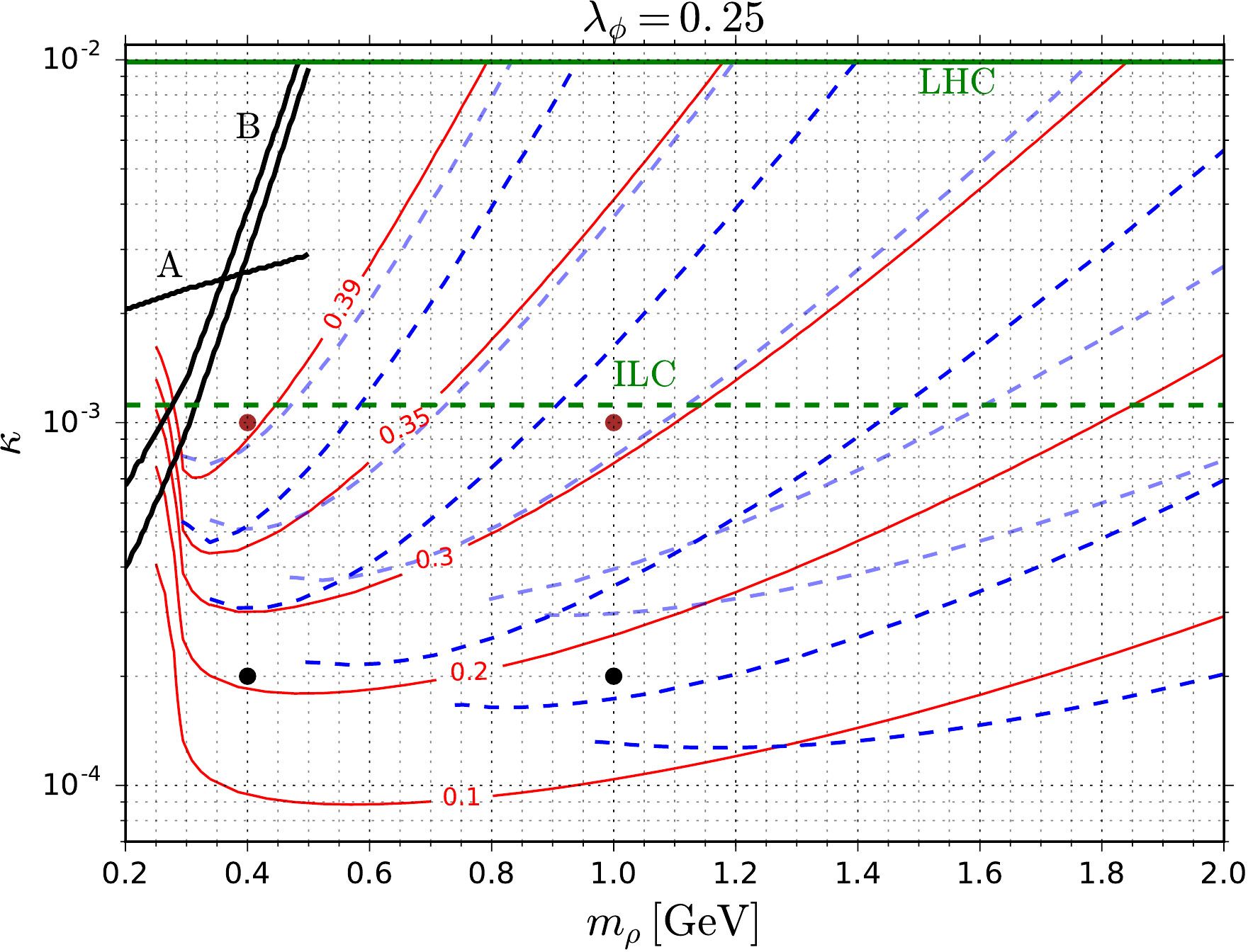}\hspace{1ex}
\includegraphics[width=0.49\textwidth]{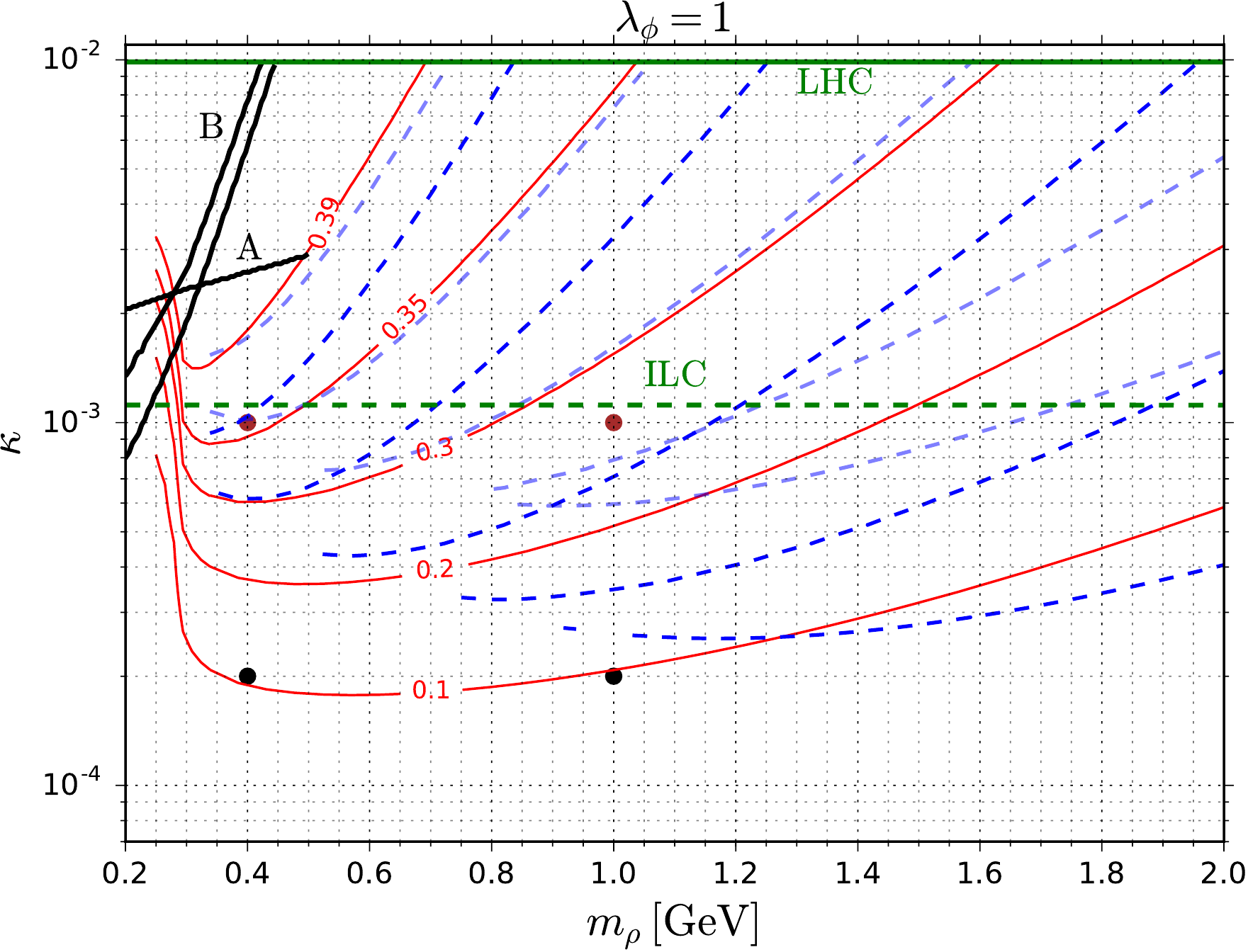}\\
\vspace{3ex}
\includegraphics[width=0.49\textwidth]{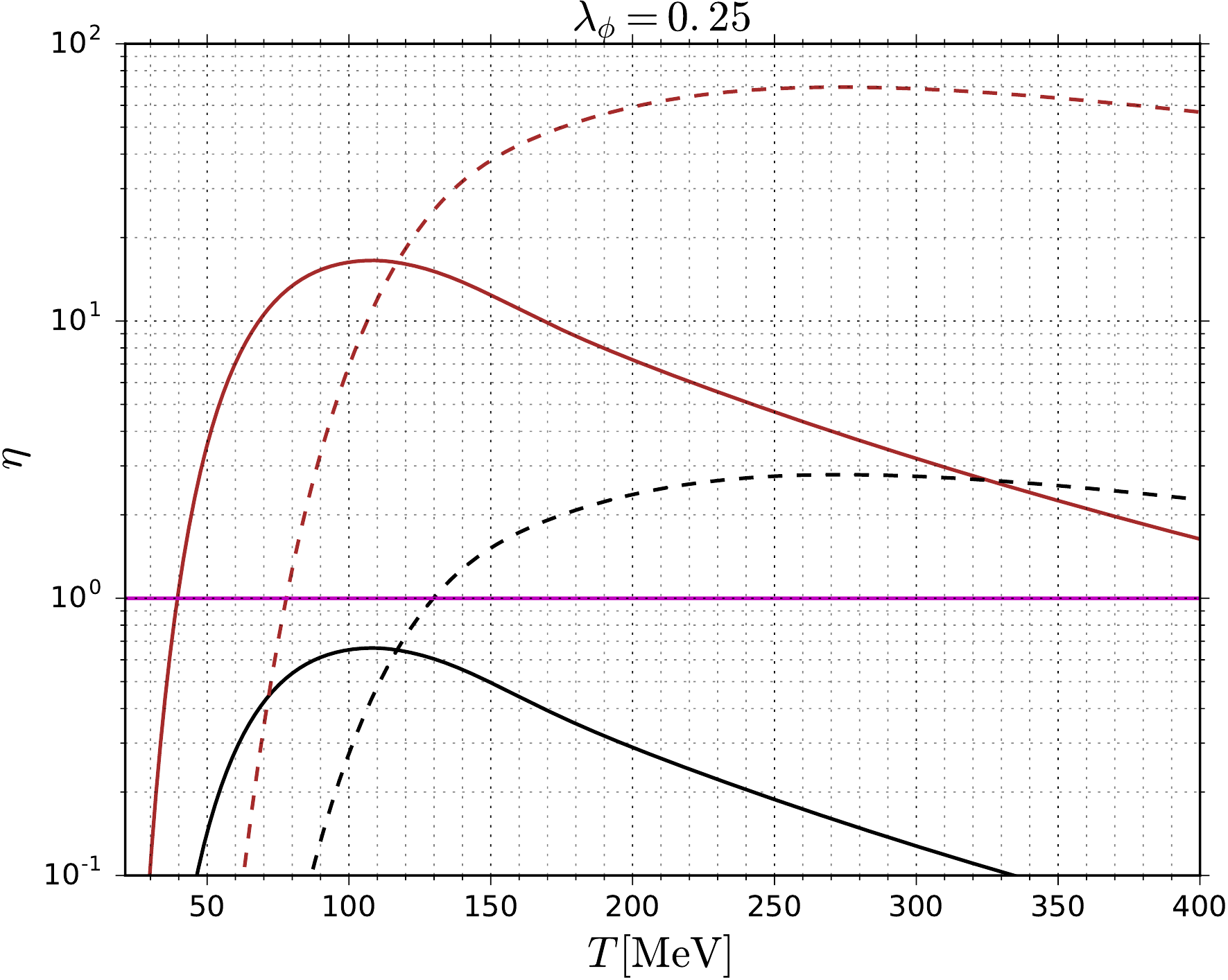}
\includegraphics[width=0.49\textwidth]{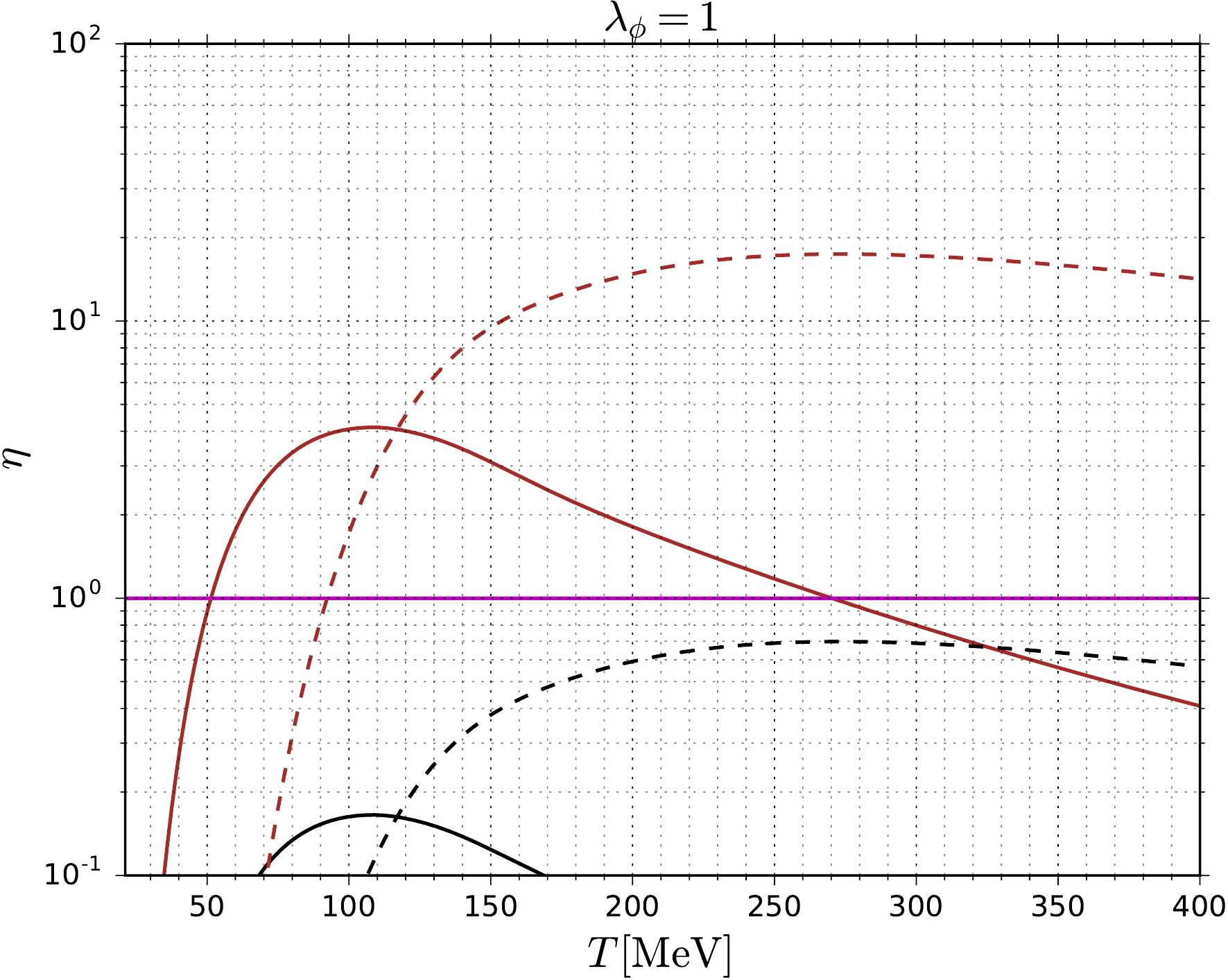}
\vspace{1ex}
\caption{Top: contour lines of constant $\Delta N_{\rm eff}$ in
  ($m_\rho$,~$\kappa$) plane for $\lambda_\phi=0.25$ (left panel) and
  $\lambda_\phi=1$ (right panel) obtained with eq.~\eqref{freeze_app}
  in Maxwell-Boltzmann approximation (dark/light blue dashed lines include
  annihilation into muons and pions/muons only) and eq.~\eqref{BE_zx} with
  full particles statistics (red lines~-- see details in sec.~\ref{subs:W_Beq}).
  Black lines correspond to astrophysical bounds~\cite{supernowe}: for
  Goldstone boson free-streaming out of the proto-neutron stars (A) and for
  energy loss due to exotic species in SN 1987A in two different approximations
  (B). Green lines denote current (LHC~-- eq.~\eqref{W_kappa_LHC}) and future
  (ILC for $B_{\rm inv}=0.3\%$) limit on invisible Higgs decay. We did not
  include effects from the CDM sector. Bottom: $\eta(T)$ dependence
  (see eq.~\eqref{freeze_app_MB}) for large dots marked in the upper panels~--
  brown (black) lines correspond to $\kappa=10^{-3}\; (2\cdot 10^{-4})$,
  whereas solid (dashed) to $m_\rho=0.4\; (1)$~GeV.
  The freeze-out process for parameters corresponding to the black dots
  for $\kappa=2\cdot 10^{-4}$ is presented in more detail in
  Fig.~\ref{fig:xBar_Y1}. }
\label{fig:freeze_app}
\end{figure}
\begin{equation}
\mathcal{C}
\frac{x_f^4}{\sqrt{g_*(x_f)}} {\rm K}_1\left(\frac{x_f\, m_\rho}{m_{\mu^\pm}}\right)
=1\,,
\label{freeze_app_MB}
\end{equation}
where
\begin{equation}
\label{freeze_app_MB_C}
\mathcal{C}\equiv
\frac{\kappa^2}{\lambda_\phi}
\frac{\sqrt{45}\,\zeta(3)}{32 \pi^{5/2}}
\frac{m_\rho^5 M_{\rm Pl}}{m_h^4 m_{\mu^\pm}^2}
\left[
\left(1-\frac{4m_{\mu^\pm}^2}{m_\rho^2}\right)^{3/2}+
\frac{1}{27}
\frac{m_\rho^2}{m_{\mu^\pm}^2}
\left(1+\frac{11}{2}\frac{m_\pi^2}{m_\rho^2}\right)^{2}
\left(1-\frac{4m_{\mu^\pm}^2}{m_\rho^2}\right)^{1/2}
\right]\,.
\end{equation}
In our numerical calculations we use the number of the effective degrees
of freedom, $g_*(x)$, as given in~\cite{Drees:g(x)}.
In the upper panels of Fig.~\ref{fig:freeze_app} we plotted $\Delta N_{\rm eff}$
obtained in the instantaneous freeze-out approximation, $Y=Y_{\rm eq}(x_f)$,
as a function of $m_\rho$ and $\kappa$ (blue lines), using
definition~\eqref{deltaNeff_def_exp} and the freeze-out temperature
calculated from eq.~\eqref{freeze_app_MB}.
The results are compared to those obtained by solving the Boltzmann
equation (red lines)~-- see sec.~\ref{subs:W_Beq}. In almost all cases the 
instantaneous freeze-out approximation overestimates $\Delta N_{\rm eff}$
for a given values of $\kappa$ and $\lambda_\phi$ (dark blue lines are
below corresponding red lines). The best agreement between the instantaneous
freeze-out approximation and the full result is achieved for
$\Delta N_{\rm eff}$ close to 0.1, although in that region the 
instantaneous freeze-out approximation works only for $m_\rho\gtrsim 1$~GeV.
Note also that neglecting the $\sigma$ annihilation into pions
(given by eq.~\eqref{W_M2_pions_delta}), while
keeping only annihilation into muons \eqref{W_M2_mions_delta}
(light blue lines in Fig.~\ref{fig:freeze_app}),
leads to quite substantial underestimation of the resulting $\Delta N_{\rm eff}$ 
(the difference amplifies when $\kappa$ decreases).\footnote{
  One can see that for larger $\Delta N_{\rm eff}$ the freeze-out approximation
  gives better agreement with the Boltzmann equation solutions
  when only $\sigma$ annihilation into $\mu^\pm$ is taken into account,
  than including also annihilation into $\pi^\pm$ and $\pi^0$.
  In particular, for $\Delta N_{\rm eff}=0.3$ one can observe a quite accurate
  coincidence in almost whole range of $m_\rho$. But this is just an
  example of situations when effects resulting from two wrong
  assumptions/approximations partially cancel out, leading to a final
  result being close to the correct one.}

It is worth emphasizing that the instantaneous
freeze-out approximation breaks down for
very small $\kappa$ and $m_\rho\lesssim 1$~GeV. Thus, exploration of this
part of the parameter space requires a more accurate approach, which we will
discuss in the next section. Let us only add that the above-mentioned problem is
related to the stiffness of condition~\eqref{freeze_app}.
As we can see in the lower panels of Fig.~\ref{fig:freeze_app}, for small
enough $\kappa$ the $\eta(T)$ function never reaches~1.

The current experimental limits, shown in Fig.~\ref{fig:freeze_app}, exclude
$\kappa\gtrsim 10^{-2}$ and the region of small $m_\rho$ with
$\kappa\gtrsim 10^{-3}$. Note also that we conservatively do not consider
here the CDM sector, which could (depending on the dark matter mass) constrain the parameter space even further~\cite{1312.2547},
\cite{supernowe}.

\subsection{Solutions of the Boltzmann equation
  \label{subs:W_Beq}}

In this subsection we compare the results obtained from solutions of the 
Boltzmann equation (with statistics of incoming and outgoing
particles taken into account) with those obtained with the approximate
methods introduced in subsection~\ref{subs:app} (which include some effects
of incoming particles statistics) and with the instantaneous
freeze-out approximation discussed in the previous subsection.

Regions of small $\kappa$, where the last approximation breaks down, are
especially important from the viewpoint of near future experiments that will
be able to probe $\Delta N_{\rm eff}$ with accuracy even better than $0.03$.
Let us start the discussion by considering two sample points from the
parameter space with small $\kappa$: two lower (black) dots in the upper
left panel of Fig.~\ref{fig:freeze_app}.
\begin{figure}[t!]
\center
\includegraphics[width=0.49\textwidth]{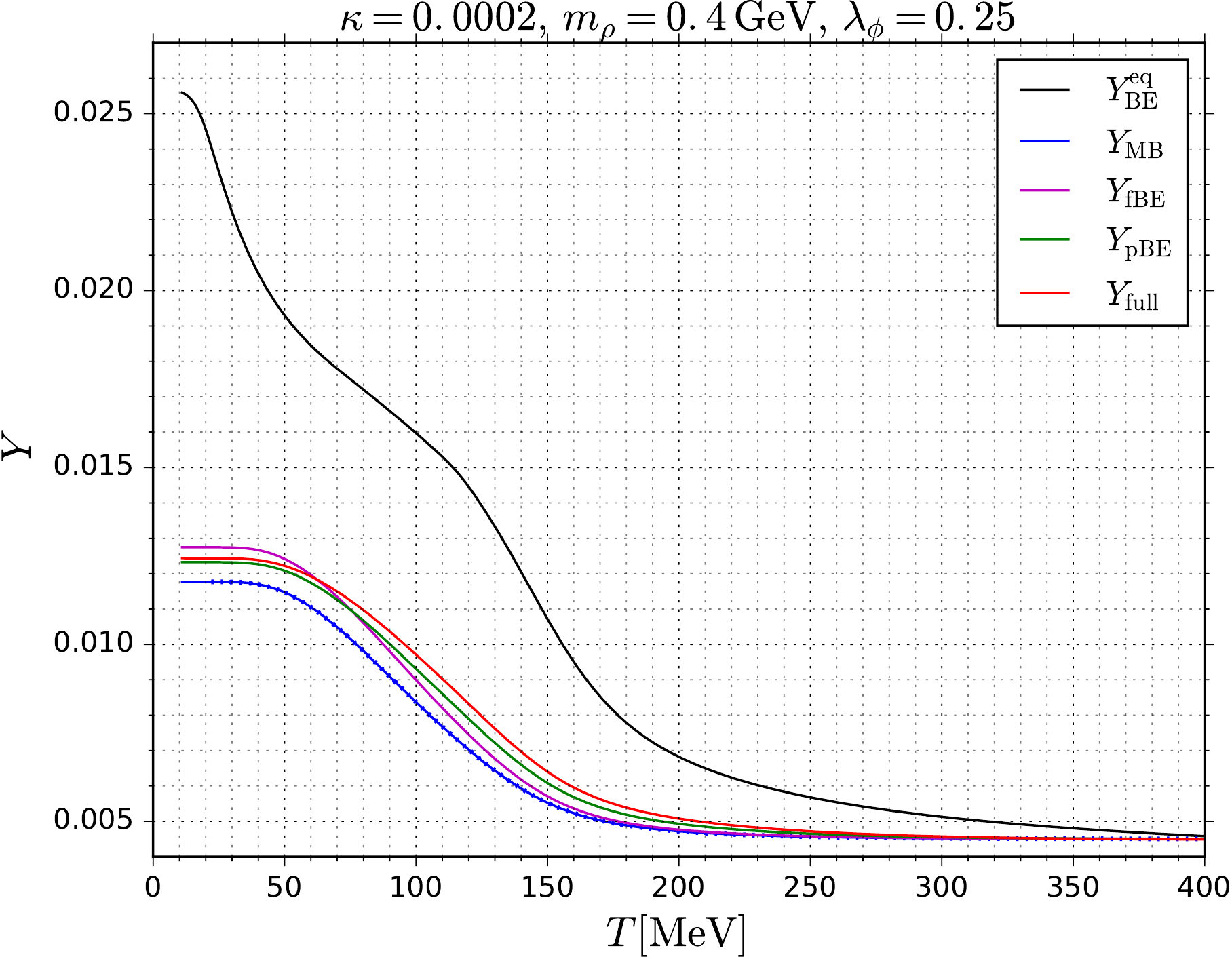}\hspace{1ex}
\includegraphics[width=0.49\textwidth]{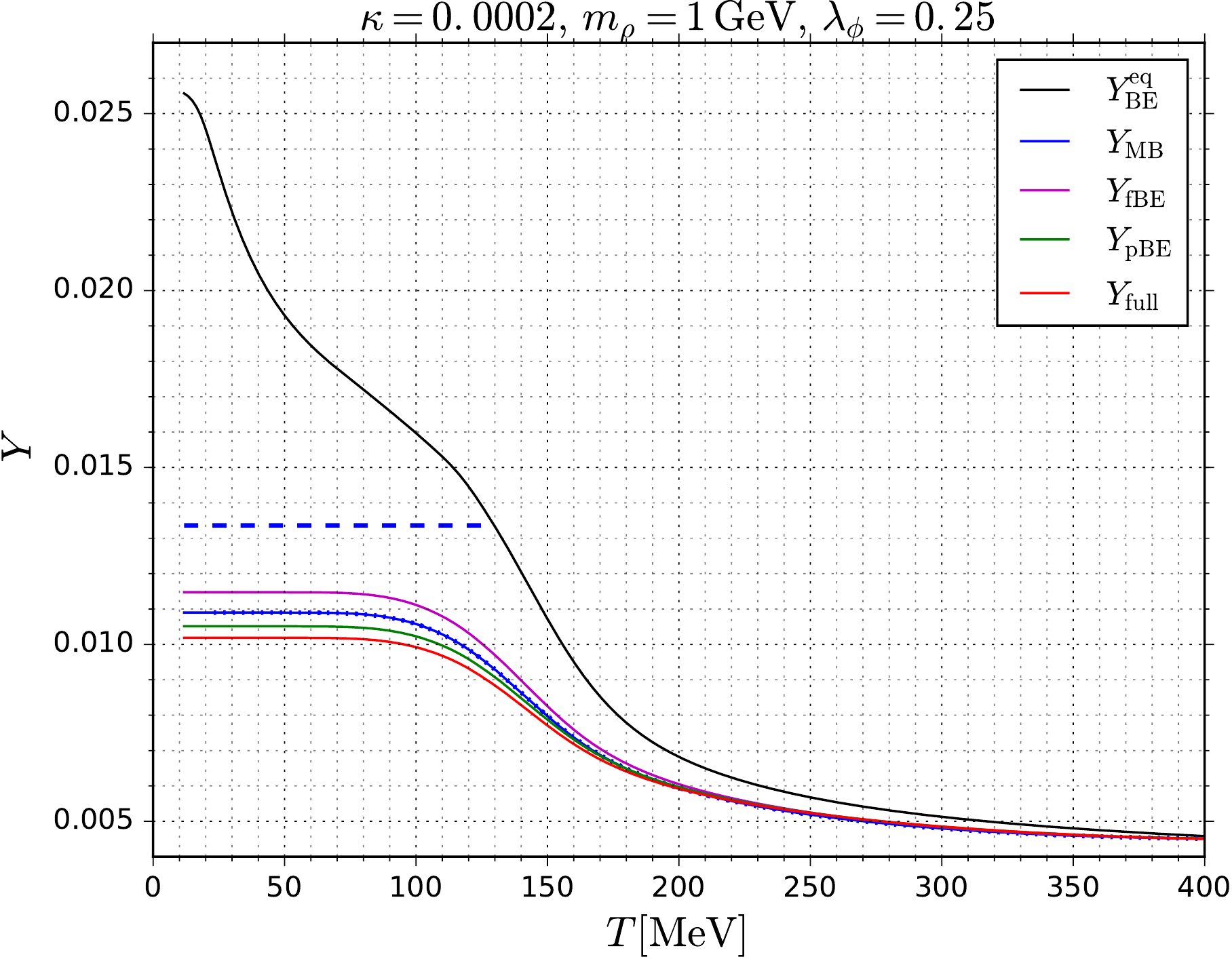}\\
\vspace{2ex}
\includegraphics[width=0.49\textwidth]{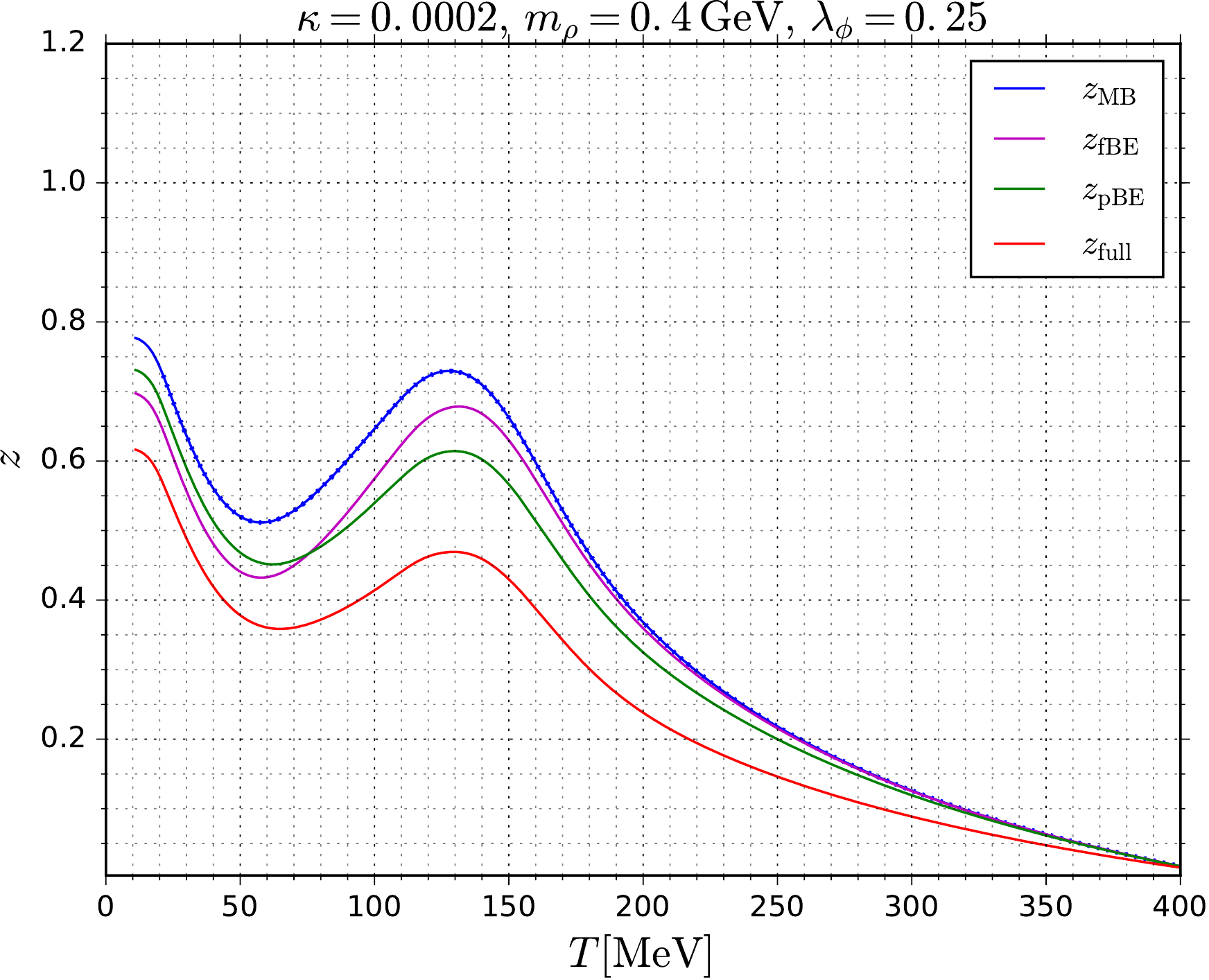}\hspace{1ex}
\includegraphics[width=0.49\textwidth]{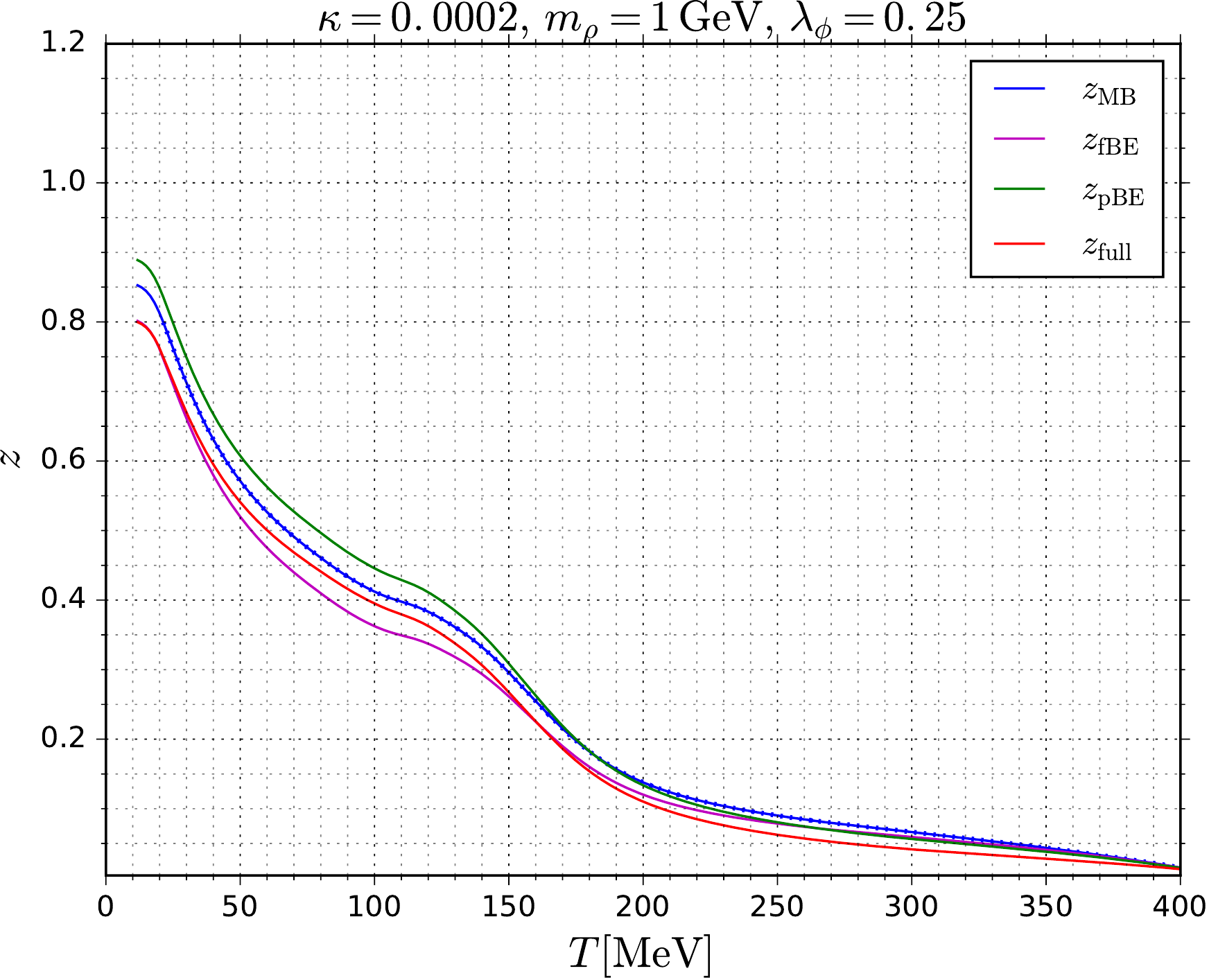}
\caption{Yield $Y$ (upper panels) and pseudopotential
  $z$ (lower panels) of Goldstone boson $\sigma$
  as functions of temperature for $\lambda_\phi=0.25$,
  $\kappa=2\cdot10^{-4}$ and for two values of  $m_\rho$~--
  see black dots in the upper left
  panel of Fig.~\ref{fig:freeze_app}. Line colors (corresponding to the methods
  used) are described in the legends. Dashed blue line in the upper right panel
  denotes the instantaneous freeze-out approximation result  (see Fig.~\ref{fig:freeze_app}).
  }
\label{fig:xBar_Y1}
\end{figure}
In Fig.~\ref{fig:xBar_Y1} we show corresponding evolutions of $Y(T)$ and $z(T)$
for all three approximations described in section~\ref{subs:app} as well as
for the most accurate of our methods. Line colors i.e.~blue, violet, green
and red correspond to the pure MB approximation (point 1.\ under
eq.~\eqref{sigmav_MB}), fractional inclusion of incoming particles
statistics (point 2.\ under eq.~\eqref{sigmav_MB}), partial inclusion
of incoming particles statistics (point 3.\ under eq.~\eqref{sigmav_pBEpFD})
and full inclusion of particles statistics (eq.~\eqref{BE_zx}), 
respectively. As we have already seen in Fig.~\ref{fig:freeze_app},
the instantaneous freeze-out approximation
(dashed blue line in the upper right panel in Fig.~\ref{fig:xBar_Y1})
overestimates the relic density, as compared to the solutions of the Boltzmann
equation (with full or any approximate inclusion of statistics
effects). Such behavior is related to the convexity of $g_{*}(T)$ function,
which for $T\sim 100\div200$~MeV and $30\div50$~MeV is characterized by strong
variability. As a result, it is harder for $\sigma$ particles to follow the
equilibrium density i.e.\ larger $\kappa$ is required as compared to the
instantaneous freeze-out approximation,
where the effect from $g_{*}(T)$ is point-wise. We checked that
the instantaneous freeze-out approximation gives results closer to those
obtained by integrating the Boltzmann equation when $g_{*}(T)$ during the
$\sigma$ freeze-out changes relatively mildly (e.g.\ for $T\gtrsim 200$~MeV).
However, the accuracy obtained with the instantaneous freeze-out approximation
is typically worse than the accuracy of other approximate methods
discussed in this work. One can also include the backreaction of $g_\sigma(x)$
on $g_{*}(x)$, however we checked numerically that this effect is negligible
in the whole range of the analyzed parameter space.

It is also worth noting in the left panels of Fig.~\ref{fig:xBar_Y1}
that $\sigma$ freezes in for $T\sim 60\div130$~MeV.
For such temperatures the pseudopotential $z(T)$ decreases with time,
which is related to the fact that the annihilation cross section for small
$m_\rho$ reaches its maximum in this temperature range (see also $\eta(T)$
dependence in the lower panels of Fig.~\ref{fig:freeze_app}).

\begin{figure}
\center
\includegraphics[width=0.49\textwidth]{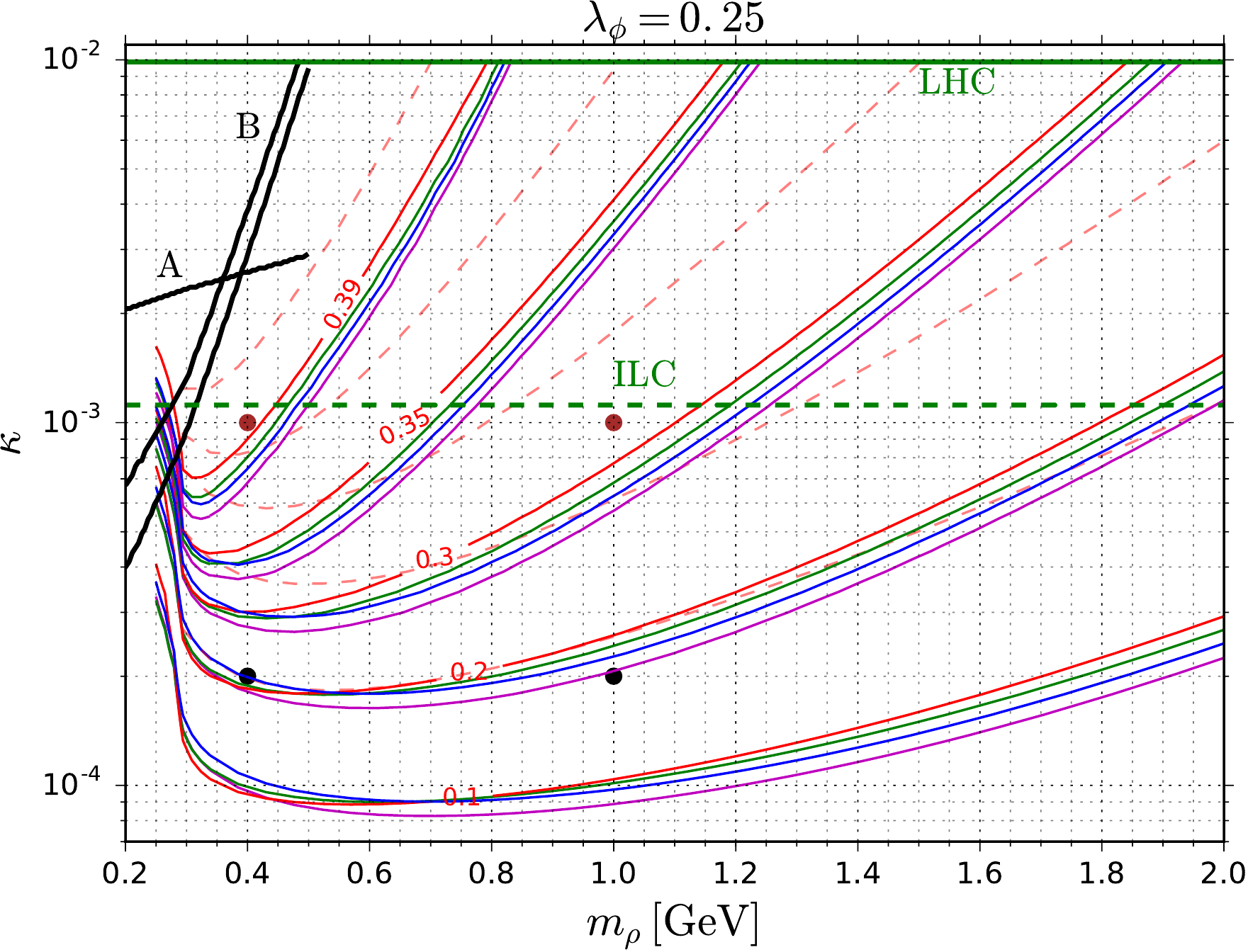}\hspace{1ex}
\includegraphics[width=0.49\textwidth]{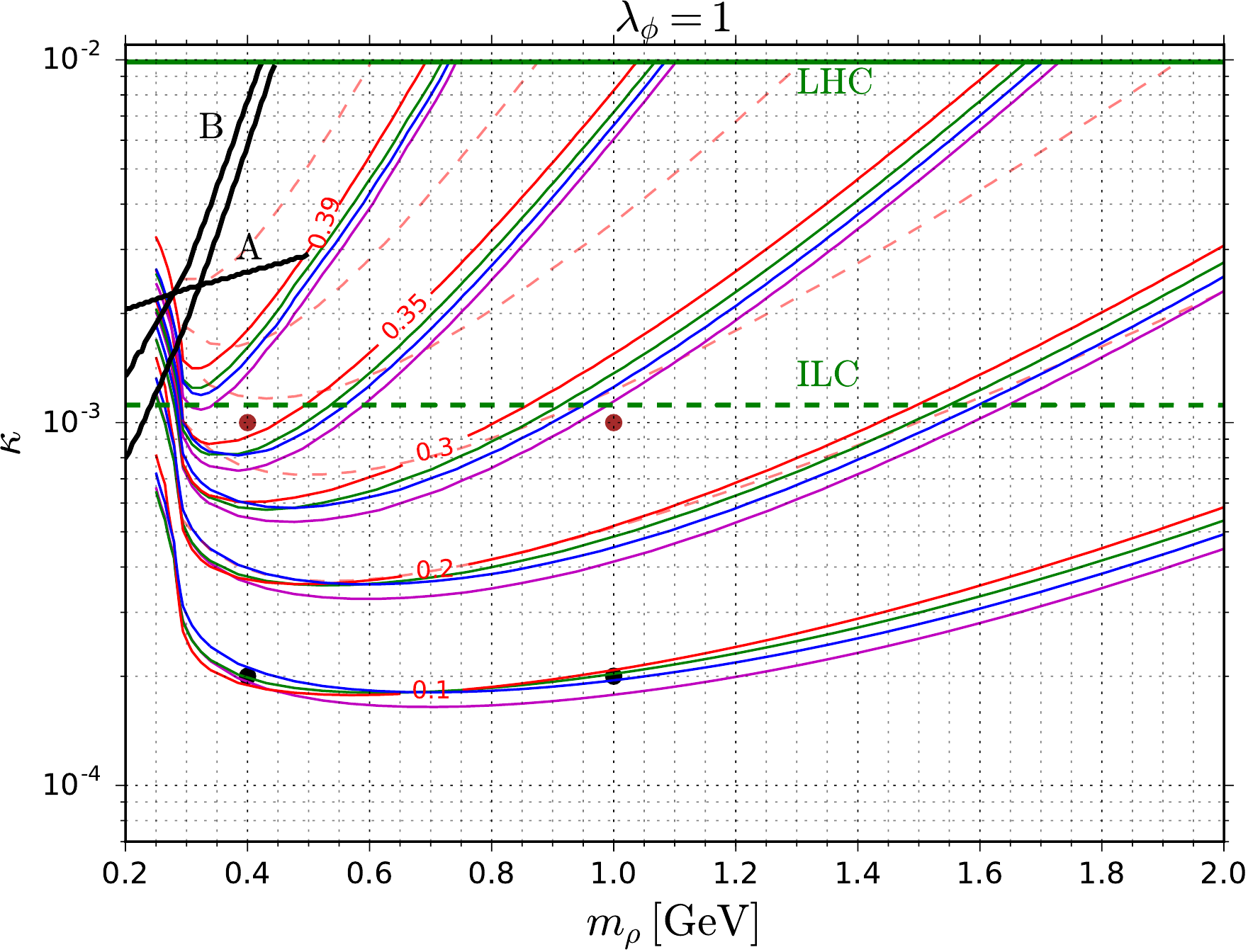}\\
\vspace{1ex}
\caption{As top panels in Fig.~\ref{fig:freeze_app} but for
  solutions of the Boltzmann equation with and without approximations
  defined in subsection \ref{subs:app}. Colors for continuous lines are
  described in Fig.~\ref{fig:xBar_Y1}. Red dashed lines correspond to
  calculation with the pions contribution neglected.}
\label{fig:m1_kappa_Neff}
\end{figure}

Having discussed general features of the lines presented in
Fig.~\ref{fig:xBar_Y1}, let us now take a closer look at the differences
between approximations in solving the Boltzmann equation, described in
sec.~\ref{subs:app}. We checked both analytically and
numerically that in the range of analyzed parameter space the
phase space integral~\eqref{Dphi_int_cos_Wdelta} may be expressed for different
configurations of incoming/outgoing particles statistics
as\footnote{
  In four letter subscripts (BEFD, FDBE, BEBE and FDFD) the two first (last)
  letters denote the statistics of incoming (outgoing) particles.}
\begin{align}
\label{DPhi_ACC=-2_fBE}
\mathscr{D}\Phi_{\rm fBE} &\approx \mathscr{D}\Phi_{\rm MB}\times
\zetai\,,
&
\mathscr{D}\Phi_{\rm fFD} &\approx \mathscr{D}\Phi_{\rm MB}\times
\zetai\,,
\\
\label{DPhi_ACC=-2_pBE}
\mathscr{D}\Phi_{\rm pBE} &\approx \mathscr{D}\Phi_{\rm MB}\times
  \frac{1}{\zetai} \coth\left(\frac{m_\rho x}{4 m_{\mu^\pm}}\right)\,,
&
\mathscr{D}\Phi_{\rm pFD} &\approx \mathscr{D}\Phi_{\rm MB}\times
\frac{1}{\zetai}
\tanh\left(\frac{m_\rho x}{4 m_{\mu^\pm}}\right)\,,
\\
\label{DPhi_ACC=-2_BEFD}
\mathscr{D}\Phi_{\rm BEFD} &\approx \mathscr{D}\Phi_{\rm MB}\,,
&
\mathscr{D}\Phi_{\rm FDBE} &\approx \mathscr{D}\Phi_{\rm MB}\,,
\\
\label{DPhi_ACC=-2_BEBE}
\mathscr{D}\Phi_{\rm BEBE} &\approx \mathscr{D}\Phi_{\rm MB}\times
\coth^2\left(\frac{m_\rho x}{4 m_{\mu^\pm}}\right)\,,
&
\mathscr{D}\Phi_{\rm FDFD} &\approx \mathscr{D}\Phi_{\rm MB}\times
\tanh^2\left(\frac{m_\rho x}{4 m_{\mu^\pm}}\right)\,,
\end{align}
where (see eq.~\eqref{sigmav_MB_Wdelta})
\begin{equation}
\label{Dphi_MB_Wdelta}
\mathscr{D}\Phi_{\rm MB}=
e^{-2z}\frac{\sum_{\rm spins}|M|^2}{2x}
\sqrt{\frac{m_\rho^2}{m_{\mu^\pm}^2}-4}\;
{\rm K}_1\left(\frac{x\,m_\rho}{m_{\mu^\pm}}\right)\,.
\end{equation}
This clearly shows that for the cases when the incoming and outgoing
particles have different statistics (BEFD or FDBE) the effects
of statistics of initial and final states cancel each other
to large extend and the MB approximation works quite well.
When the initial and final particles have the same statistics one
gets amplification (BEBE) or suppression (FDFD) with respect to
the MB approximation. 
In Weinberg's Higgs portal model one has to consider a combined
case i.e.\ BEFD (annihilation into muons) and BEBE (annihilation into pions).
The effect from the latter (factor $\coth^2\left(m_\rho x/4 m_{\mu^\pm}\right)$)
starts to be important for small $m_\rho$ and $x$ (larger $T$)
and also for small $\kappa$. The dependence on $\kappa$ follows from the
fact that $\sigma$ with smaller $\kappa$ decouples at higher temperature
i.e.~at smaller $x$. These effects can be seen
in Fig.~\ref {fig:m1_kappa_Neff},
where for small enough $m_\rho$ and $\kappa$ red curves are placed partially
under the blue ones i.e.\ in order to get a given value of $\Delta N_{\rm eff}$
smaller $\kappa$ is needed when using the Boltzmann equation with full
statistics effects as compared to the MB approximation.

One can also observe from Fig.~\ref{fig:m1_kappa_Neff} that
for a given value of $\kappa$ and $m_\rho$ in most cases 
$\Delta N_{\rm eff}^{\rm pBE}<\Delta N_{\rm eff}^{\rm MB}<\Delta N_{\rm eff}^{\rm fBE}$
and relative differences between approximations decrease when $\kappa$
increases. Only for small enough $\kappa$ and $m_\rho$ we have
$\Delta N_{\rm eff}^{\rm pBE}>\Delta N_{\rm eff}^{\rm MB}$.
These relations can be easily understood with the help of  
eqs.~\eqref{DPhi_ACC=-2_fBE} and \eqref{DPhi_ACC=-2_pBE}. 
Moreover, for the most part of the parameter space (i.e.\ except the
region with sufficiently small $m_\rho$
and $\kappa$, where pions statistics matters) full inclusion of incoming and
outgoing statistics results in smaller $\Delta N_{\rm eff}$ than the
above-mentioned approximations. There are two effects which explain this
behavior. Firstly, from eq.~\eqref{n} one can see that for given $z$ and $x$
inclusion of incoming particles statistics  (here: BE) results in 
$n(x)$ (equivalently $Y(x)$) smaller than in the MB approximation
(see eq.~\eqref{y_and_Yeq_MB}) and the difference increases with~$z$.
Secondly, comparing eq.~\eqref{BE_zx} and~\eqref{z_MB} one can show (see
definition~\eqref{J})
that for given $z$
and $x$ coefficients multiplying two terms in the RHS of eq.~\eqref{BE_zx}
($J_3(z,x)/J_2(z,x)$ and $1/J_2(z,x)$ respectively) are smaller than those
in eq.~\eqref{z_MB}. First term in the RHS of eq.~\eqref{BE_zx} dominates
for small $z$ (when $Y(x)$ traces $Y_{\rm eq}(x)$) which causes weaker change
in $|{\rm d}z/{\rm d}x|$ as compared to the MB case. When the second term
starts to be important (for larger $z$) the coefficient $1/J_2(z,x)$
effectively weakens the annihilation strength (given by $\tilde{S}_I(z,x)$)
leading to faster $|{\rm d}z/{\rm d}x|$ change with respect to other
approximations. Both effects are visible in Fig.~\ref{fig:xBar_Y1}.

Contrary to naive expectations the MB approximation gives better accuracy
than the fBE one, which includes the effect from incoming
particles statistics only in $Y_{\rm eq}$. In order to obtain more precise
results, it is necessary to take into account the effects
from statistics also in the calculation of the thermal average
of the appropriate cross section (pBE).
Let us add that taking into account
$\sigma$ annihilation into muons only (see Fig.~\ref{fig:freeze_app} and
Fig.~\ref{fig:m1_kappa_Neff}) leads to sizable discrepancies with respect to
the case when full annihilation is considered.
The only exception holds for $m_\rho\leq 2m_\pi$ (i.e.\ when annihilation into
pions does not take place)~-- one can observe that continuous lines rapidly
change slope and red lines converge towards dashed ones.

Relative differences between our approximations may reach
$\Delta N_{\rm eff}\sim 0.05$ and
more.\footnote{
  This effect for a Goldstone boson with just one degree of freedom is of
  similar magnitude as the summed effect of kinetic non equilibrium during 
  decoupling for six neutrino degrees of freedom in the SM 
  ($N_{\rm eff}^{\rm SM}\approx 3.046$).}
Thus, statistics of both  incoming and outgoing particles are
relevant, especially for moderate values of $\kappa$ and $m_\rho$. Minimal
value of 
$\Delta N_{\rm eff}$ obtained in the scan ($\sim 0.06$) is related to the
starting moment for the calculation i.e.\ $T_{\rm start}=400$~MeV, however for
larger $T$ $Y_{\rm eq}(T)$ does not change significantly. Therefore, if $\sigma$
freezes out during or after the QCD phase transition, it shall give 
contribution to $N_{\rm eff}$ measurable by near future experiments.
On the other hand, maximal $\Delta N_{\rm eff}$
equals approximately 0.5 and is achieved for large $\kappa$ and $m_\rho$ near
the $h_1$ resonance (see eq.~\eqref{narrow_res}), which is not preferred due
to collider and astrophysical bounds.
It is worth mentioning that the widely studied scenario with
$\Delta N_{\rm eff}\sim 0.39$ for moderate values of $\lambda_\phi$ can be
probed in major parts of the parameter space by the ILC (and new generation
of CMB satellite experiments). However, the region with smaller
$\lambda_\phi$ may be more
challenging in this context as the lines of constant $\Delta N_{\rm eff}$
move towards smaller $\kappa$, becoming harder to be probed.

\begin{figure}
\center
\includegraphics[width=0.49\textwidth]{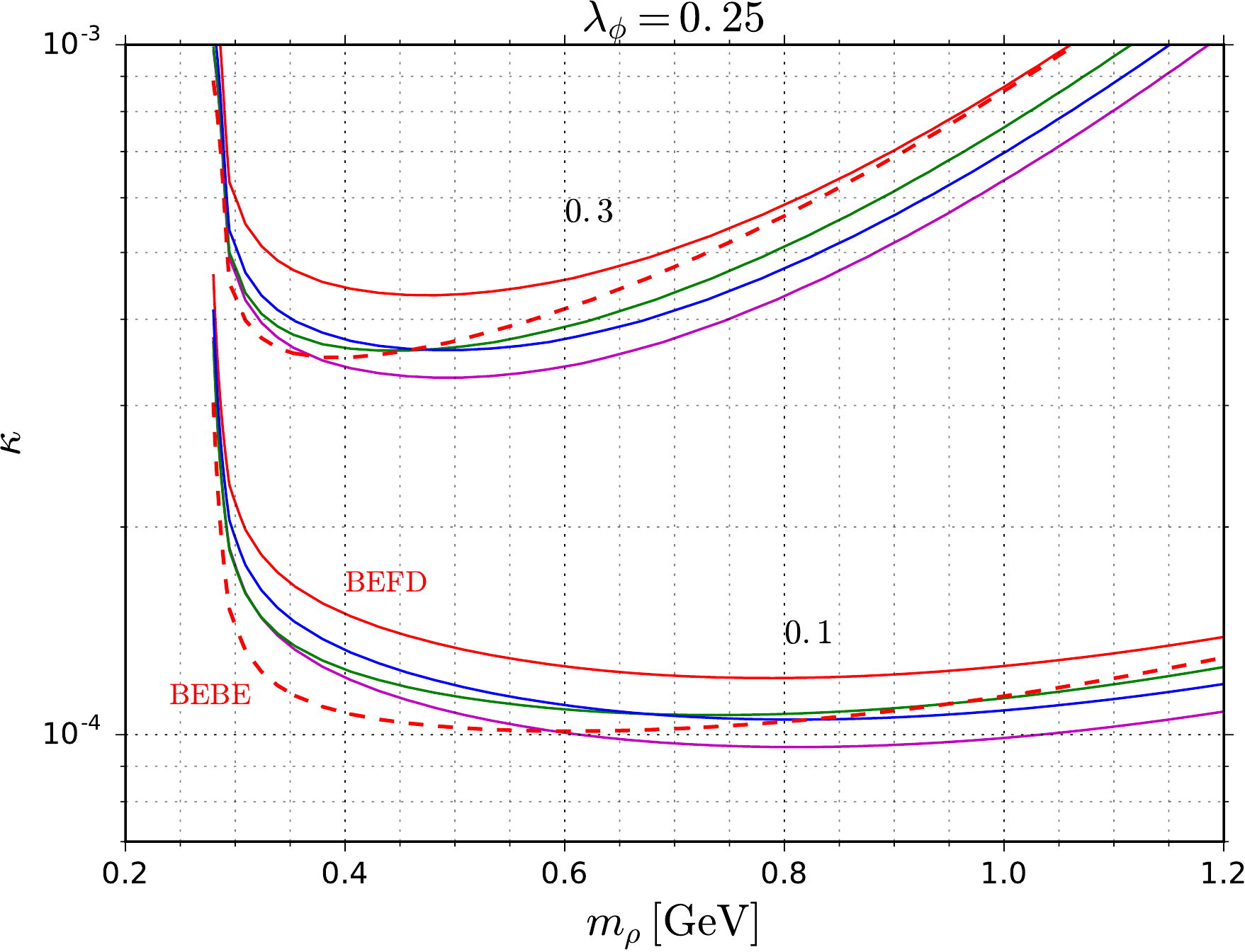}\hspace{1ex}
\includegraphics[width=0.49\textwidth]{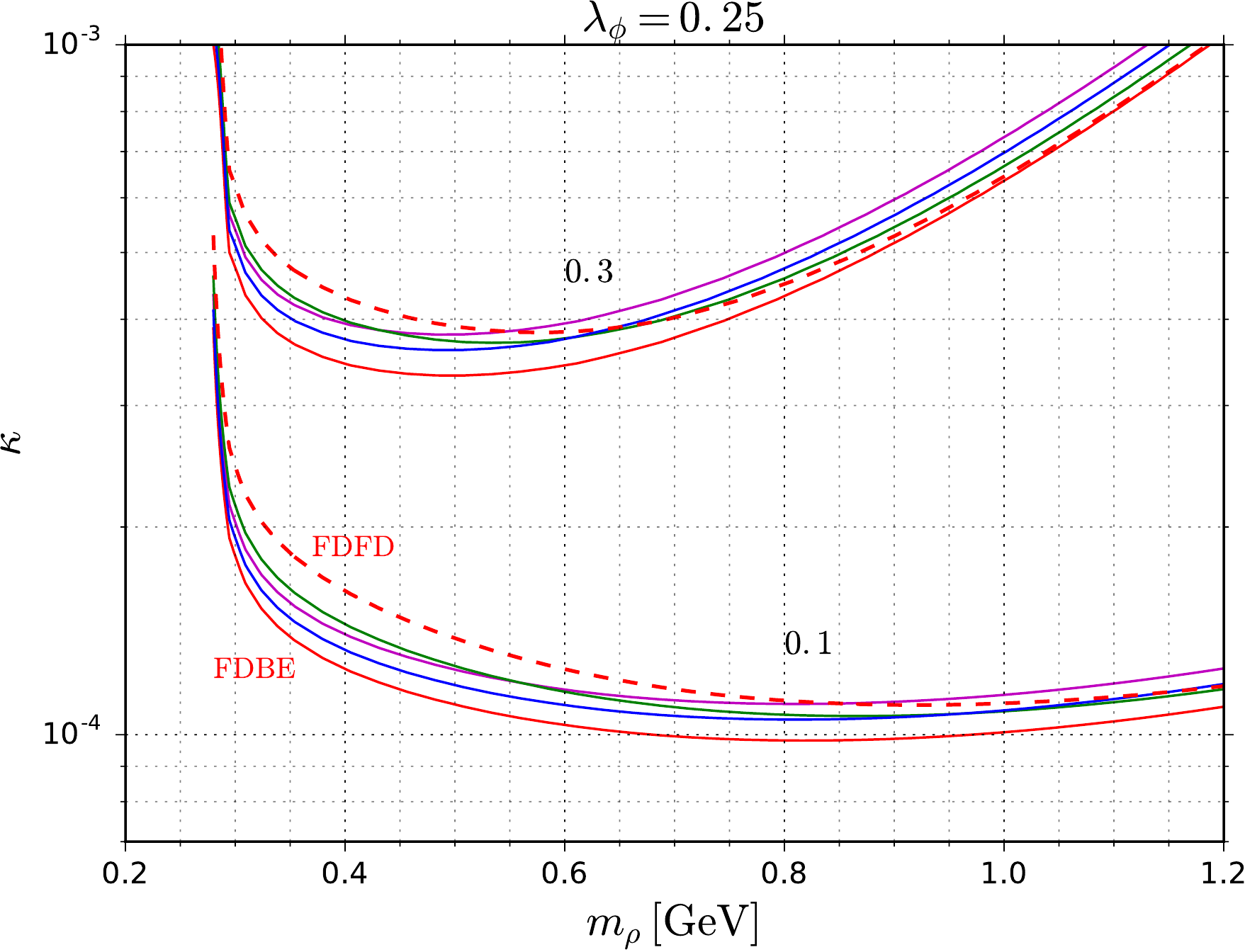}\\
\vspace{1ex}
\caption{Contour lines of constant $\Delta N_{\rm eff}$ (top lines: 0.3, bottom lines: 0.1) in ($m_\rho$,~$\kappa$) plane for $\lambda_\phi=0.25$ and bosons/fermions annihilation (left/right panel). Continuous (dashed) red lines correspond to mixed (homogeneous) statistics~-- see eqs.~\eqref{DPhi_ACC=-2_BEFD} and \eqref{DPhi_ACC=-2_BEBE} e.g.\ BEBE case refers to $\sigma$ annihilation into pions in Weinberg's Higgs portal model. Colors for other lines (MB, fBE/fFD, pBE/pFD) are the same as in the previous plots. }
\label{fig:m1_kappa_Neff_stat}
\end{figure}

Let us finish this section with the discussion of differences between
possible combinations of incoming and outgoing particles statistics
(see eqs.~\eqref{DPhi_ACC=-2_BEFD}
and \eqref{DPhi_ACC=-2_BEBE}). In Fig.~\ref{fig:m1_kappa_Neff_stat} we show
similar plots to those presented in Fig.~\ref{fig:m1_kappa_Neff} but
for\footnote{
  Asymptotic behavior for larger $m_\rho$ and $\kappa$ resembles that from
  the previous plots.
}
$m_\rho\leq 1.2$~GeV, $\kappa\leq 10^{-3}$ and taking into account only
(dominant) annihilation into pions (eq.~\eqref{W_M2_pions_delta}). 
Then, BEBE case corresponds to Weinberg's Higgs portal model with annihilation
into muons neglected, while the others  correspond to toy models with
statistics of incoming and/or outgoing particles changed to the FD one.
For mixed statistics BEFD/FDBE continuous red curves are placed above/below
those that were obtained including the effects from incoming particles
statistics only i.e.\ fBE/fBE and pBE/pFD (as well as from the MB
approximation). For such mixed cases the effects of statistics of
incoming and outgoing particles approximately cancel out in $\mathscr{D}\Phi$
(see eq.~\eqref{DPhi_ACC=-2_BEFD}). The main reason for smaller/bigger
values of $\Delta N_{\rm eff}$ in such mixed cases, especially when compared
to the results of the MB approximation, is the presence of the statistics
factor $\si=\pm1$ in eqs.~\eqref{J} and \eqref{n} (as discussed
below eq.~\eqref{Dphi_MB_Wdelta}).
As one can see (and what can be observed also in
Fig.~\ref{fig:m1_kappa_Neff_stat}) for homogeneous statistics (BEBE and FDFD)
the additional factors   
($\coth^2\left(m_\rho x/4 m_{\mu^\pm}\right)$ and 
$\tanh^2\left(m_\rho x/4 m_{\mu^\pm}\right)$) in  $\mathscr{D}\Phi$
given by \eqref{DPhi_ACC=-2_BEBE}) 
are important for small exchanged particle mass (here $m_\rho$)
and coupling (here $\kappa$). 
The effects of outgoing particle statistics decrease
with $m_\rho$ and $\kappa$ and eventually become negligible.

\section{Conclusions
  \label{sec:conclusions}}

We have investigated the problem of calculating the relic abundance 
of Dark Radiation which freezes out before the SM neutrinos decoupling.
We used the Boltzmann equation for relativistic particles with their
statistics taken into account. This method was compared to the 
instantaneous freeze-out approximation and to some approximate methods 
in which statistics of DR particles is included only in a limited way 
or completely ignored. As an interesting illustration of all these methods we 
analyzed in some detail the relic density of DR -- measured by the 
change of the effective number of neutrino species $\Delta N_{\rm eff}$ 
-- in the Weinberg's Higgs portal model. The main results are as follows:
\begin{itemize} 
\item
  The popular instantaneous freeze-out approximation can not be
  applied for small values of $m_\rho$ and $\kappa$ for which the
  Boltzmann equation must be used.
\item
  In most of the remaining regions of the parameter space the instantaneous
  freeze-out approximation overestimates $\Delta N_{\rm eff}$. The main reason
  is convexity of $g_*(T)$ which in this simple method is not taken
  into account. The instantaneous freeze-out approximation gives best results
  for $\Delta N_{\rm eff}\sim 0.1$ and $m_\rho\gtrsim 1$~GeV. The resonant
  exchange of the scalar $\rho$ (for $m_\rho$ up to a few hundreds MeV)
  also plays an important role.
\item 
  When calculating the annihilation cross section of DR particles
  (Goldstone bosons $\sigma$) it is crucial to include muons and pions
  as the final states. Pions are quite often ignored which may lead
  to underestimation of $\Delta N_{\rm eff}$ by as much as 0.1.
\item
  Not taking (fully) into account the statistics of DR particles and
  particles into which DR annihilates may change the obtained values of
  $\Delta N_{\rm eff}$ by up to about 0.05.
  Contrary to naive expectation, in some parts of the parameter space
  ignoring the effects of statistics (MB approximation) may give better
  prediction for $\Delta N_{\rm eff}$ than inclusion of only some of the
  effects -- those in evaluation of $Y_{\rm eq}(T)$ -- due to the DR
  statistics (fBE approximation). Inclusion of more effects from the DR
  statistics (pBE approximation) leads to more accurate results than those
  obtained using the simplest MB approximation.
\end{itemize}

The present experimental data leave quite substantial uncertainty 
in determining the value of $\Delta N_{\rm eff}$. One may expect that 
results of near future experiments will lead to much better 
determination of $\Delta N_{\rm eff}$ and will allow to test scenario considered in this work. In such a case it will be 
important not only to use the statistics of incoming and outgoing particles 
in the appropriate Boltzmann equation but maybe also to take into account 
the effects of deviations from kinetic equilibrium. This may be 
especially important for cases with resonant exchange of $\rho$ 
when DR particles $\sigma$ may decouple kinetically before thermal 
decoupling. This might happen because in elastic scattering particle $\rho$
is exchanged in the t channel for which there is no resonant
enhancement~\cite{Binder:2017rgn}.
Another potentially important issue is the relation between the DR 
and CDM sectors. We assumed in our analysis that CDM particles are
so heavy that they do not influence the DR properties. However,
it may occur, especially when future more precise experimental results
are available, that the case of lighter CDM should be considered
simultaneously with DR.

\section*{Acknowledgements}

This work has been partially supported by National Science Centre,
Poland, under research grants DEC-2014/15/B/ST2/02157
and DEC-2012/04/A/ST2/00099.
MO acknowledges partial support from National Science Centre, Poland, 
grant DEC-2016/23/G/ST2/04301.
PS acknowledges support from National Science Centre, Poland, 
grant DEC-2015/19/N/ST2/01697.


\end{document}